\documentclass[12pt]{article}

\input epsf

\font\eulerm=eusm10 scaled \magstep1
\font\eulerb=eusb10 scaled \magstep2

\def\W{\mbox{\eulerm W}}
\def\WW{\mbox{\eulerb W}}

\def\i{\mbox{i}}
\def\d{\mbox{d}}
\def\Im{\mbox{Im}}
\def\Re{\mbox{Re}}

\title{ \vspace*{2.5cm} \bf \Large
        Finite width effects and gauge cancellations \\
        in W- and Z-boson production in framework of \\
        modified perturbation theory \vspace*{5 mm}}
\author{M. L. Nekrasov
\smallskip  \\ {\small
Institute for High Energy Physics, 142280 Protvino, Russia}}

\date{}

\begin{document}

\maketitle

\begin{abstract}

{The processes of production and subsequent decay of W- and
Z-bosons in $e^+ e^-$ collisions are considered in a recently
proposed modified perturbation theory (PT), based on a direct
expansion of probabilities instead of amplitudes. In such an
approach the non-integrable singularities in the phase space,
which are intrinsic in the conventional PT, appear as
singularities in the coupling constant (with subsequent
compensation by the decay factors of unstable particles). In the
present paper the systematic investigation of the modified PT is
carried out. The results are compared with the results of the
conventional approach, based on calculation of the amplitude with
Dyson resummation. A solution to the problem of the loss of
one-loop PT order in the resonance region is found. On the basis
of this solution the proof of gauge cancellations in any order of
the modified PT is given. A simple generalization of the
fermion-loop scheme is proposed which provides a complete
description of W-pair production in next-to-leading order
approximation.}

\end{abstract}

\newpage

\section{Introduction}
\label{sec:1}

In many applications of the Standard-Model, connected with the
present and future collider experiments, one should take into
account the effects of the instability of W- and Z-bosons (as well
as of Higgs boson, top quark etc.) \cite{LEP2}. In quantum field
theory the conventional way to take into account an instability
consists in the Dyson resummation of the self-energies of unstable
particles \cite{Veltman}. This procedure avoids nonintegrable
phase-space singularities caused by the processes of production
and decay of intermediate unstable particles. However, Dyson
resummation leads to a deviation from the scheme of fixed-order
calculations in the framework of perturbation theory (PT). In
gauge theories this results in the violation of the Ward
identities (WI) and loss of gauge invariance
\cite{LEP2-1,NuovoCim,B-P}. This fact leads to loss of the control
of the high-energy behavior of the theory and to the emergence of
large errors in the description of particular processes.

In the case of single Z-boson production (LEP1) and within the
precision defined by one-loop corrections to the vertex functions,
the problem of the gauge invariance was solved ad hoc (in fact,
the mentioned precision implies the approximation of
next-to-leading order, NLO). The most consistent scheme of
calculations was described in \cite{B-P}, where only the
gauge-invariant contributions to the self-energy were Dyson
resummed, while the gauge-dependent ones were considered by the
conventional PT. As a result, the amplitude could be presented as
a product of gauge-independent factors, two vertex and one
resonant. Nevertheless, this result is not universal. Anyway, now
it is not clear whether this holds within the next order of
precision determined by two-loop corrections to the vertex
functions.

In the case of a pair production of unstable particles (LEP2) the
amplitude cannot be presented in the framework of Dyson
resummation in a completely gauge invariant form\footnote{It
should be noted that there is an alternative approach called the
pole scheme, described in \cite{Stuart} and then elaborated in
numerous papers. The gauge invariance in this scheme is initially
maintained, but, unfortunately, an algorithm for the evaluation of
corrections is not developed completely (see \cite{Doubl-Pole} and
discussion in Sect.\ref{sec:8}).}. The hopes, nevertheless, for
any further progress in such calculations are usually connected
with the rather general idea of the determination of the minimal
set of Feynman diagrams that are necessary for compensating the
gauge violation by the Dyson resummed self-energies. For this
purpose within the NLO approximation the fermion-loop scheme was
proposed \cite{Argyres,Fermion-loop}. Nevertheless, the bosonic
corrections were not taken into consideration in this approach. In
order to solve this problem a generalization of the fermion-loop
scheme was proposed \cite{Background}, defined in terms of the
formalism of the background-field method. It solved the problem of
the bosonic corrections and, moreover, it remained in force also
beyond the one-loop approximation. However, this approach could
not solve another problem, which was in the fermion-loop scheme,
too. This problem concerns the incompleteness of the description
within the declared precision of description because of the loss
of one-loop order of PT in the resonance region
\cite{Dittmaier,Acta}.

Let us consider in more detail the latter phenomenon. The point is
that the denominator of unstable-particle propagator in the
resonance region, $p^2 - M^2 = O(g^2)$, is of order $O(g^2)$, but
not $O(1)$. So, in this region the Dyson resummed one-loop
self-energy actually makes a contribution in the leading-order,
but not in NLO. Therefore, in order to complete the NLO
approximation the two-loop correction to the self-energy should be
Dyson-resummed.

However, the Dyson resummation of the two-loop correction, made
without taking into consideration the two-loop corrections to the
vertex functions, hardly leaves a chance to maintain WI.
Consideration of all two-loop corrections maybe solves the
problem. Nevertheless, this solution is certainly impractical
\cite{Dittmaier,Acta}. Besides, the question remains unclear how
the two-loop corrections to the vertex functions, which in fact
contribute in NNLO, can compensate the gauge violation that occurs
in NLO. So, anyway the problem of the loss of one-loop PT order
has to be solved in a practical fashion.

The present paper proposes the solution to this problem in the
framework of the modified PT \cite{Base}. Its basic idea is the
expansion of direct probabilities in powers of the coupling
constant instead of amplitudes. (The amplitudes prior to
calculation of the probabilities are considered to be full.) Such
an order of operations, taken together with ideas of the theory of
generalized functions \cite{Schwartz,G-Sh}, allows one to trace
the fundamental connection between the origin of the phase-space
non-integrable singularity and the loss of one-loop PT order in
the resonance region. (In fact this loss manifests itself as the
emergence of an extra singularity in the coupling constant instead
of the phase-space singularity.) Moreover, while considering the
probability the contribution of the Dyson resummed two-loop
correction (in the amplitude) may be reproduced within the given
precision in the form of an additive anomalous term. Owing to the
additivity of this term, it becomes possible to give an
independent proof of the gauge cancellations in the probability.
It is worth mentioning that this result does not mean that the
inclusion of two-loop corrections only to denominators of
``unstable'' propagators (without the inclusion to vertices) does
not lead to violation of WI in the amplitude. This means that
contributions that violate WI in the amplitude turn out to be
beyond the given precision in the probability. By its nature this
phenomenon is connected with the effect of changing of the order
of individual contributions in the probability due to the
emergence of a singularity in the coupling constant instead of the
phase-space singularities.

Below we elaborate the above-stated ideas in any order of the
modified PT and give a general proof of the gauge cancellations in
the probability within the given precision. Notice, that the
latter outcome was practically anticipated in pioneering work of
\cite{Base}, presenting the modified PT. However the reasoning of
this work in the part that concerns the gauge cancellations was
not complete. Indeed, it was based on a comparison with results of
the conventional approach, but it overlooked the problem of losing
one-loop PT order in the resonance region. Besides, it omitted the
problem of the difference between remainders of the expansions of
the amplitude and the corresponding probability. Nevertheless,
proceeding from only the amplitude it is impossible to estimate in
a mathematically correct way the remainder of the expansion of the
probability, since in the resonance region the expansion of
amplitude faces the ambiguity of the expression 0/0.

In order to investigate in detail the above-mentioned problem we
first perform the systematic study of the modified PT approach.
Then we proceed with applying the results in the theory of
electroweak interactions. In general case we find a practical way
to keep a fixed precision without violating gauge cancellations in
the framework of the background-field method. Within NLO we also
find the solution in the usual formalism, basing ourselves on the
results of the fermion-loop scheme.

This paper is organized as follows. In Sect.~\ref{sec:2} a
statement of the problem is discussed by a simplified example of a
single unstable particle production. The basic formulas of the
modified PT are derived in Sect.\ref{sec:3} (as a whole, the
content of this section follows \cite{Base}). The properties of
the expansion of the unstable propagator squared are studied in
Sect.\ref{sec:4}. Section \ref{sec:5} discusses briefly the
soft-photon problem. In Sect.\ref{sec:6} the general proof is
given of the gauge cancellations in processes mediated by unstable
particle production. Section \ref{sec:7} is devoted to
construction of the generalization of the fermion-loop scheme in
NLO approximation. In Sect.\ref{sec:8} the results are discussed.

\section{Unstable propagator in AO: \\ statement of the problem}
\label{sec:2}

In this section we consider the structure of the denominator of an
unstable propagator. We reject, for a moment, all factors in the
numerator. Then let us consider the propagator in the following
form:
\begin{equation}\label{1}
\Delta (\alpha ;\tau )={1 \over {M^2-p^2-\Sigma }}\,\,\,=\,\,\,{1
\over {\tau -\alpha \kern 1pt h(\tau )-\i \alpha f(\tau )}}\,\,.
\end{equation}
Here $\alpha = g^2\!\left/(4\pi)\right.$ is the coupling constant
squared, $\tau = M^2 - p^2$ is the kinematic variable, $M$ and $p$
are the mass and momentum of the unstable particle (the mass
squared is considered without the conventional $-\i 0$), $\Sigma$
is the renormalized self-energy,\footnote{In the case of vector
bosons the self-energy includes two structures, $g_{\mu\nu}$ and
$p_{\mu} p_{\nu}$. Formula (1) represents the contribution of the
first structure only. The effect of presence of the $p_{\mu}
p_{\nu}$-structure will be discussed latter on.} $\alpha h$ and
$\alpha \!f$ are its real and imaginary parts with the extracted
factor $\alpha$. Here we do not fix the scheme of UV
renormalization because the results will not depend on it (see
below). By definition, the property of instability means that $f
\not= 0$ in some neighborhood of the point $\tau = 0$. Owing to
causality we assume that $f > 0$ in this neighborhood. In what
follows we assume that the size of this neighborhood is of the
order of $O(\alpha)$ and that it contains a solution to the
equation $\tau - \alpha h(\tau) = 0$. Generally speaking, the
function $h(\tau)$ may be nonzero at $\tau = 0$.

The probability of the process of the production and decay of an
unstable particle is defined by an integral over some region of
the kinematic variable of the propagator squared ${\W}(\alpha
;\tau)$, multiplied by some weight function,
\begin{equation}\label{2}
P(\alpha )=\int {\d\tau \;\varphi (\tau )\;}{\W}(\alpha ;\tau ),
\end{equation}
\begin{equation}\label{3}
{\W}(\alpha ;\tau )\equiv \left| {\Delta (\tau )} \right|^{\,2}={1
\over {\kern 1pt \left[ {\tau -\alpha \,h(\tau )\kern 1pt }
\right]^{\kern 1pt 2}+\alpha ^2f^2(\tau )}}.
\end{equation}
The weight function $\varphi(\tau)$ stands for the complementary
part of the unitarity diagram with respect to the given propagator
squared. It includes the photons radiation from the initial and
final charged states. In the general case $\varphi(\tau)$ includes
also hardware factors of experimental devices.

By virtue of the property of $f \not= 0$ the function $\W
(\alpha;\tau)$ is finite and, therefore, is integrable in a
neighborhood of $\tau = 0$. However, in the limit $\alpha \to 0$
the non-integrable singularity $1\!\left/{\tau^2}\right.$ appears
in $\W (\alpha;\tau)$. This fact means the impossibility of direct
application of the conventional PT in the case of processes
mediated by unstable particles. Nevertheless, the expansion in the
coupling constant does exist in the probability $P(\alpha)$.

Actually, the origin of non-integrable singularity means that the
\emph{result of integration} of $\W (\alpha;\tau)$ with a weight
function $\varphi(\tau)$ includes a singularity in $\alpha$ at
$\alpha \to 0$. If one extracts this singularity, then the
expansion of the remaining integral becomes possible. In the case
of a power-like singularity, this expansion will be a Laurent
expansion, but not a Taylor one. (Let us note that the weight
function $\varphi(\tau)$ actually depends on the parameter
$\alpha$, including it, in particular, as a factor. Therefore, the
expansion of the integral ultimately may take a Taylor form.
However, a priori the kind of the singularity is not known. So in
order to study the problem, we first consider $\varphi(\tau)$ as a
function independent from $\alpha$. We also assume that
$\varphi(\tau)$ is a rather smooth function and generally is non
zero at $\tau = 0$.)

Let us study the kind of singularity of the integral. Since the
singularity originates from integration over small $\tau$, we may
keep in the functions $h(\tau)$ and $f(\tau)$ only their leading
terms of the asymptotic expansion at $\tau \to 0$, i.e.
approximate them by $h_0 = h(0)$ and $f_0 = f(0)$. As a result, up
to inessential corrections (which will be calculated later on), we
get the approximation as follows:
\begin{equation}\label{4}
{\W}(\alpha ;\tau )\cong {1 \over {[\tau -\alpha \,h_0]^2 +
\alpha^2f_0^2}}\;.
\end{equation}
By virtue of the homogeneity one can deduce from (4) that the
integral of $\W (\alpha;\tau)$ with the weight function
$\varphi(\tau)$ has a singularity of $1\!\left/ \alpha \right.$.
Indeed, let us divide the range of integration into $|\tau| >
\mbox{const}\!\times\!\alpha$ and $|\tau| <
\mbox{const}\!\times\!\alpha$ with large enough ``const''. The
integration over the first range gives a finite contribution,
while the integration over the second range gives the
above-mentioned singularity (this may be verified by a change of
the integration variable). Moreover, the coefficient at the
singularity is proportional to $f^{-1}_0$ and does not depend on
$h_0$. In fact, $f_0$ may be included into the normalization of
$\alpha$, whereas $h_0$ does not make contribution in the leading
order. Indeed, setting $h_0 = 0$ does not lead to a singularity in
$\tau$. For similar reasons the weight function $\varphi(\tau)$
gives a contribution to the leading-order term as a trivial factor
$\varphi(0)$.

So, despite the fact that the expansion of the integrand is an
incorrect operation, the expansion of the \emph{result of
integration} is worthwhile. Moreover, some properties of such
expansion may be determined before the actual calculation of the
integral. For systematic determination of the properties there is
a special method called the asymptotic operation, AO
\cite{IJMP,PNP,Homo}. Its key point is the transition to an
extended interpretation of the integrand as a product of a kernel
of \emph{generalized} function on a \emph{test} function
\cite{Schwartz,G-Sh}. In fact this means that the integral is
interpreted as a continuous linear functional on a test function.
When the integral is well defined (\emph{before} expanding the
integrand) the above-mentioned generalization does not lead to any
modification. However, after the formal expansion of the integrand
the new interpretation allows one to give meaning to the
non-integrable terms of the expansion.

Thus, the problem of the expansion of the integrand may be solved
basically within the method of generalized functions. Next, it is
necessary to consider the asymptotic properties of the expansion.
For this purpose an ambiguity of the extension of the
interpretation of nonintegrable functions in the integrand is
used. Generally, it is well-known (e.g. from experience of the UV
renormalizations) that elimination of divergences may be
accompanied by ambiguities. When an integral is determined by the
method of generalized functions, the ambiguities are described by
means of so-called counterterms, which are proportional to the
delta-function or its derivatives located strictly in the point of
the non-integrable singularity.\footnote{ Let us emphasize that
the introduction of counterterms is a general place in the theory
of generalized functions (see e.g. \cite{G-Sh}). Actually this
idea was used by N.N.Bogoliubov \cite{R,B-Sh} for establishing the
R-operation. In the AO context the term ``counterterms'' was
introduced \cite{IJMP,PNP} in order to emphasize the analogy with
the theory of UV renormalizations.} In the AO framework the
coefficients at counterterms are unambiguously fixed by the
requirement of the reconstruction of the result which should be
obtained from an expansion of the initial integral. (Let us recall
that before expansion of the integrand the integral was well
defined and there were no ambiguities in it.) Moreover, AO gives a
practical recipe of the calculation of these coefficients in each
order of the expansion prior to a calculation of the integral. The
resulting counterterms contain complete information about the
terms which are singular in the parameter of the expansion.
Simultaneously the counterterms may also contain some non-singular
contributions which correct the asymptotic property of the
expansion.

In the above example the counterterm that describes the leading
term of the asymptotic expansion of $\W (\alpha;\tau)$ is
$c/(\alpha f_0)\times\delta(\tau)$ with $c$ is some numerical
factor. In the given case the value of $c$, as well as the
appearance of this counterterm, follows from the formula which is
well-known in the theory of generalized functions,
\begin{equation}\label{5}
\lim_{\alpha \to 0} \frac{\alpha}{\tau ^2+\alpha^2} =
\pi\delta(\tau ) .
\end{equation}
Thanks to this formula the expansion of $\W (\alpha;\tau)$ up to
an $O(1)$ correction is defined unambiguously and turns out to be
the delta-function only. Up to the $O(\alpha)$ correction the
expansion is non-trivial. Its most general form is
\begin{equation}\label{6}
{\W}(\alpha ;\tau )={\pi  \mathord{\left/{\vphantom {\pi {\left(
{\alpha f_0} \right)\,}}} \right. \kern-\nulldelimiterspace}
{\left( {\alpha f_0} \right)\,}}\delta (\tau ) +\left[ {\,{{1}
\mathord{\left/ {\vphantom {{1} {\tau ^2}}} \right. \kern-
\nulldelimiterspace} {\tau ^2}}} \right]+c_0\,\delta (\tau
)+c_1(-)\,\delta '(\tau )+O(\alpha ).
\end{equation}
Here $\left[ {\,{{1} \mathord{\left/ {\vphantom {{1} {\tau ^2}}}
\right. \kern- \nulldelimiterspace} {\tau ^2}}} \right]$ is the
generalized function defined with some prescription. (The most
common but not mandatory prescription is the principal value.
Later on we omit the square brackets.) The last two terms in (6)
are counterterms that correct the contributions of the first two
terms.

In the general case the complete determination of the generalized
function $1\!\left/{\tau^2}\right.$ may be done by the following
way \cite{G-Sh}. First, one makes two subtractions in the test
function $\varphi(\tau)$ in some neighborhood of $\tau = 0$ by
replacing $\varphi(\tau)$ to $\varphi(\tau) - \varphi(0) - \tau
\varphi'(0)$. As a result, the non-integrable singularity in
$1/{\tau^2}$ becomes compensated. (In fact this is one of possible
prescriptions.) Then in order to describe the ambiguity emerging
with this subtraction, one has to add two counterterms to
$1/{\tau^2}$. One counterterm is proportional to the
delta-function and the other one is proportional to its first
derivative (both counterterms correspond to the above
subtractions). The coefficients at the counterterms must be
determined in such a way \cite{IJMP,PNP,Homo} as to guarantee the
asymptotic properties of the expansion of the integral at order
$O(1)$. Their values depend on $h$ and $f$, and on the choice of
the prescription in $1\!\left/{\tau^2}\right.$, but the sum of all
terms will not depend on the prescription.

The above process may be continued. The next term of the formal
expansion of $\W (\alpha;\tau)$ is $2\alpha h(\tau )\times {{1}
\mathord{\left/ {\vphantom {{1} {\tau ^3}}} \right.
\kern-\nulldelimiterspace} {\tau ^3}}$. For its complete
determination one needs three counterterms which include the
delta-function, its first and its second derivatives. The
coefficients at counterterms are fixed by the requirement of the
asymptotic properties of the expansion. The practical recipe of
the calculation of the coefficients is presented in the next
section.

\section{Calculation of counterterms }
\label{sec:3}

Let us show the technique of the calculation of counterterms on
the example of the AO expansion of $\W (\alpha;\tau)$ up to
$O(\alpha^2)$. Since the leading term of this expansion is of
order $O(\alpha^{\!-1})$, the mentioned precision is sufficient
for the determination of the next-to-next-to-leading order (NNLO)
approximation. Such a precision seems to be sufficient for most of
the practical applications. It is sufficient also for
understanding the properties of the AO expansion of $\W
(\alpha;\tau)$.

The general structure of the AO expansion of $\W (\alpha;\tau)$
within the considered precision is as follows:
\begin{equation}\label{7}
{\W}(\alpha ;\tau )\ =\ {1 \over {\tau ^2}}+{{2h(\tau )} \over
{\tau ^3}}\alpha +E(\tau )+O(\alpha ^2)\; .
\end{equation}
Here the first two terms represent the result of the formal
expansion of $\W (\alpha;\tau)$. For definiteness, we treat the
poles with respect to $\tau$ in the sense of principal value.
Below we recall two equivalent definitions of the principal value:
\begin{equation}\label{8}
V\!P{{1 \over {\tau ^n}}} = {1 \over 2}\left[ {{1 \over {(\tau\!
+\!\i 0)^n}}\!+\!{1 \over {(\tau\!-\!\i 0)^n}}}
\right]={{(-)^{n-1}} \over {(n\!-\!1)! }}\,{{\d^n} \over {\d\tau
^n}}\ln \left| {\tau \kern 1pt } \right|.
\end{equation}
The derivatives in the last expression are understood in the sense
of generalized functions, i.e. they must be swit\-ched to the test
function via formal integration by parts.

The quantity $E(\tau)$ in (7) represents the sum of counterterms.
In this case they are proportional to the delta-function, its
first and its second derivatives:
\begin{equation}\label{9}
E(\tau )=\ \sum\limits_{n\,=\,0}^2 {{{(-)^nc_n} \over {n!
}}\,}\delta ^{(n)}(\tau )  .
\end{equation}
In what follows we assume that $h(\tau)$ and $f(\tau)$ together
with their second derivatives are regular functions in some
neighborhood of $\tau = 0$.\footnote{ If the unstable particle is
able to interact with massless particles (photons), this property
does not occur. Nevertheless, supplying the photons with mass we
restore the analyticity inside a neighborhood defined by the
generated mass gap. This is enough for our purposes (see
discussion in Sect.~\ref{sec:5}).} At first, for simplicity we
suppose that the functions $h$ and $f$ include one-loop
contributions only. A generalization to the multi-loop case is
considered below in this section.

We wish to define a procedure for the determination of the
coefficients $c_n$, $n = 0,1,2$, for an arbitrary $\varphi(\tau)$
function, which decreases rapidly enough at infinity. This
procedure should provide the following:
\begin{equation}\label{10}
\int\limits_{-\infty }^{+\infty } {\d\tau }\,\varphi (\tau )\left[
{{\W}(\alpha ;\tau )\!-\!{1 \over {\tau ^2}}\!-\!{{2h(\tau )}
\over {\tau ^3}}\,\alpha - E(\tau )} \right]=O(\alpha ^2) .
\end{equation}
Note, that for the solution of this problem we do not have to know
all information about the function $h(\tau)$ in the third term in
square brackets. We need only have knowledge about three terms of
its asymptotic expansion,
\begin{equation}\label{11}
h(\tau )=h_0+\tau \,h'_0+(\tau^2/2)\,h''_0+o(\tau ^2) .
\end{equation}
This property follows from the fact that the remainder $o(\tau^2)$
cancels the nonintegrable singularity $1\!\left/ \tau^3 \right.$
in (10).

Now let us substitute (9) into (10) and, following \cite{IJMP},
present the test function $\varphi(\tau)$ as a linear combination
of three basis functions $\varphi_{n}(\tau)$, $n = 0,1,2$, such
that $\varphi_{n}^{(k)}(0) = \delta_{n}^{k}$, where
$\delta_{n}^{k}$ is the Kronecker symbol. As a result we get
\begin{equation}\label{12}
c_n=\!\int\limits_{-\infty }^{+\infty } {\d\tau }\,\varphi_n(\tau
)\!\left[ {{\W}(\alpha ;\tau )\!-\!{1 \over {\tau
^2}}\!-\!{{2h(\tau )} \over {\tau ^3}}\,\alpha } \right] +
O(\alpha ^2) .
\end{equation}
It is obvious from (12) that within the given precision the
coefficients $c_n$ do not depend on the choice of test functions.
Indeed, any other test functions $\widetilde \varphi_{n}(\tau)$
possessing the same property, $\widetilde \varphi_{n}^{(k)}(0) =
\delta_{n}^{k}$, lead to the coefficients $\widetilde c_n$ instead
of $c_n$. However, the difference $\widetilde c_n - c_n$ amounts
to a quantity of order $O(\alpha^2)$, because the difference
between the corresponding integrals is controlled by the test
function $\widetilde \varphi_{n}(\tau) - \varphi_{n}(\tau)$, which
equals zero at $\tau = 0$ together with its first and second
derivatives. It is easy to see that the integral in formula (12)
with such a weight function is of order $O(\alpha^2)$. Due to this
property, without loss of generality one may choose the test
functions to be the step-like ones. Namely, we may set
$\varphi_{n}(\tau) = \tau^n \times \theta(|\tau| < \Lambda)$.
Then, with $n=0,1,2$, we have
\begin{equation}\label{13}
c_n=\int\limits_{-\Lambda }^{+\Lambda } {\d\tau }\,\tau ^n\left[
{{\W}(\alpha ;\tau )\!-\!{1 \over {\tau ^2}}\!-\!{{2h(\tau )}
\over {\tau ^3}}\alpha } \right] + O(\alpha ^2) .
\end{equation}

Formula (13) basically solves the problem stated above. However,
it is still too complicated for practical usage since, generally
speaking, through $\W (\alpha; \tau)$ it contains the dependence
on the unknowns functions $h(\tau)$ and $f(\tau)$. Moreover, (13)
contains a lot of superfluous information because the integral in
r.h.s. includes contributions beyond the given precision. In
particular, the dependence on the cutoff parameter $\Lambda$ is of
this kind.

The problem may be solved by homogenization procedure \cite{Homo},
i.e. by the specific transformations in the integ\-rand. Namely,
let us do the substitutions $\tau \to \xi\tau$, $\alpha \to
\xi\alpha$. Then the result we expand in powers of $\xi$, and in
the end we set $\xi = 1$. Each term of this (secondary) expansion
is proved to be a homogeneous function of $\tau$ and $\alpha$. So
it leads to a strictly definite contribution in the power in
$\alpha$ to the integral, which now may be considered without the
cutoff. The first term $c_n^{(0)}$ of this expansion gives the
leading-order contribution to the coefficient $c_n$,
\begin{equation}\label{14}
c_n^{(0)}=\!\int\limits_{-\infty }^{+\infty } {\d\tau
}\,\tau^n\!\left[ {{1 \over {(\tau -\alpha h_0)^2+\alpha ^2f_0^2}}
\!-\! {1 \over {\tau ^2}} \!-\! {{2h_0} \over {\tau ^3}}\alpha }
\right] .
\end{equation}
The next term of the expansion of homogenization gives the
correction term $c_n^{(1)}$, etc. On counting of powers we have
$c_n^{(0)} \sim \alpha^{n-1}$, $c_n^{(1)} \sim \alpha^{n}$, etc.
Adding the necessary number of $c_n^{(i)}$ we get $c_n$ with the
required precision.

Let us emphasize, that the integral in (14) is convergent at
infinity. The convergence takes place also for other $c_n^{(i)}$.
Moreover, this property is a general result of the application of
the homogenization \cite{Homo}. It is worth recalling that the
singular terms $1\!\left/\tau^2\right.$ and
$1\!\left/\tau^3\right.$ in (14) are defined in the sense of
principal value. Nevertheless, the prescription may be changed (in
all above-derived formulas). As a result the coefficients
$c_n^{(i)}$ will change, but the asymptotic properties of the
expansion (7) will be conserved.

After the corresponding calculations we come to the following
result (for the first time obtained in \cite{Base}):
\begin{eqnarray}\label{15}
c_0 &=& {\pi\!\!\over {\alpha f_0}}+{{\pi \left( {h'_0f_0-\kern
1pt h_0f'_0\kern 1pt } \right)} \over {f_0^2}} \nonumber\\
&&+{{\pi\!\left( {{h'}_0^2f_0^2+\kern 1pt h_0^2{f'}_0^2+\kern 1pt
h_0h''_0f_0^2-\kern 1pt 2h_0h'_0f_0f'_0\kern 1pt -{\textstyle{1
\over 2}}h_0^2f_0f''_0-{\textstyle{1 \over 2}}\kern 1pt
f_0^3f''_0\kern 1pt } \right)} \over {f_0^3}}\,\alpha ,
\nonumber\\ c_1 &=& {{\pi \kern 1pt h_0} \over {f_0}}+{{\pi \left(
{2h_0{h'}_0f_0-h_0^2{f'}_0- f_0^2{f'}_0} \right)} \over
{f_0^2}}\,\alpha \;,\;\;\;\;\;\;\;\;\;\;c_2={{\pi \left( {h_0^2-
f_0^2} \right)} \over {f_0}}\,\alpha \,.\qquad
\end{eqnarray}
Here the subscript 0 means that the corresponding quantities are
defined at $\tau = 0$, while the superscript means we are taking
derivatives. For example, $h'_0 = {\d h(\tau)/\d \tau}\vert_{\tau
=0}$, etc.

The derived result may be written in a more compact form if the
quantities $c_n$ in (9) are understood as functions of $\tau$. In
this case, instead of (15), one can obtain
\begin{equation}\label{16}
    c_0={\pi  \over {\alpha f}}\;,\;\;\;\;\;\;c_1={{\pi \kern 1pt h}
    \over
f}\;,\;\;\;\;\;\;c_2=\alpha {{\pi \left( {h^2-f^2} \right)} \over
f}\;.
\end{equation}
Here $c_{\,0,1,2}$, $h$ and $f$ are functions of the $\tau$. The
equivalence of these two forms, (15) and (16), follows from the
relations $f(\tau )\,\delta '(\tau )=f_0\kern 1pt \delta '(\tau
)-f'_0\kern 1pt \delta (\tau )$ and $f(\tau )\,\delta ''(\tau
)=f_0\kern 1pt \delta ''(\tau )\,-\kern 1pt 2f'_0\kern 1pt \delta
'(\tau )+f''_0\,\delta (\tau )\ $.

The above-derived results may easily be extended to the case of
multi-loop contributions in $h(\tau )$ and $f(\tau )$. In this
case one should perform the conventional expansion in $\alpha$ in
formulas (15) or (16), in which $h$ and $f$ are understood with
the presence of multi-loop contributions \cite{Base}. The simplest
way to prove this statement is as follows: assuming that $h$ and
$f$ are the full functions we repeat all the same reasoning as was
done above, except that during homogenization we modify the
scaling by setting $\alpha^n \to \xi \alpha^n$ in the $n$-loop
contributions (instead of $\alpha^n \to \xi^n \alpha^n$). As a
result the higher-loop contributions become identified with the
one-loop ones, and formulas (15) and (16) are restored
automatically.

\boldmath
\section{The properties of AO expansion of $\WW (\alpha; \tau)$}
\label{sec:4} \unboldmath

Let us discuss now the properties of the AO expansion of $\W
(\alpha; \tau)$. At first we are interested in the characteristic
which will be useful in proving gauge cancellations in the
probability. It should be emphasized that a solution to this
problem is not obvious a priori due to the specificity of the
Feynman rules in the modified PT approach (see formulas (7), (9)
and (15)). Then we will examine the self-consistency properties of
the expansion, such as a non-sensitivity of the results to the
ambiguities in the definition of the gauge-bosons masses
\cite{Sirlin}, and consistency with the requirements of UV
renormalization.

We start from a rather methodological problem consisting in the
explicit demonstration of the independence of the results of an AO
expansion from the sequence of the expansion in the higher-loops.
In other words, we show that the result of the AO expansion of $\W
(\alpha;\!\tau)$ will be the same if instead of (1) one will start
from the following formula of the incomplete Dyson resummation:
\begin{equation}\label{17}
\Delta (\alpha ;\tau ) = \frac{\displaystyle 1}{\displaystyle \tau
-\alpha \,\Sigma_1 - \alpha ^2\,\Sigma _2 - \alpha ^3\,\Sigma _3 +
\,\cdots} = \frac{\displaystyle 1}{\displaystyle \tau - \alpha
\,\Sigma _1} + \frac{\displaystyle \alpha ^2\,\Sigma _2 +
\alpha^3\,\Sigma _3}{\displaystyle \left( {\tau -\alpha \,\Sigma
_1} \right)^2}+\,\cdots .
\end{equation}
Here $\alpha^n \Sigma_{n}(\tau)$ is the $n$-loop contribution to
self-energy. By squaring (17) we get the incomplete expansion for
$\W (\alpha; \tau)$, each term of which is integrable in the
conventional sense:
\begin{eqnarray}\label{18}
\matrix{ {\W}(\alpha ;\tau )={\W}_1(\alpha ;\tau )+\left[
{\left({\alpha ^2\,\Sigma _2+\alpha ^3\,\Sigma _3}
\right)\,{\W}_{11}(\alpha ;\tau) +\alpha ^4\left( {\Sigma _2}
\right)^2{\W}_{12}(\alpha ;\tau )+\mbox{h.c.}}
\right]\,\!\!\!\!\nonumber\\[-0.5\baselineskip]\hfill\cr
\qquad\qquad +\;\alpha ^4\left| {\kern 1pt \Sigma _2}
\right|^2\,{\W}_1^2(\alpha ;\tau )+O(\alpha ^2)\,.\hfill\cr}
\end{eqnarray}
Here we have introduced three new functions: $\W_{11}(\alpha;
\tau) = \W_{1}(\alpha; \tau) \Delta_{1}(\alpha; \tau)$, \
$\W_{12}(\alpha; \tau) = \W_{1}(\alpha; \tau)
\Delta_{1}^{2}(\alpha; \tau)$, and $\W_{1}^{2}(\alpha; \tau) =
\W_{1}(\alpha; \tau) \W_{1}(\alpha; \tau)$, where $\W_{1}(\alpha;
\tau)$ and $\Delta_{1}(\alpha; \tau)$ are the same functions as
$\W(\alpha; \tau)$ and $\Delta(\alpha; \tau)$, but with only
one-loop self-energies Dyson resummed. By considering these new
functions as generalized functions, each term in (18) may be in
turn completely AO expanded. Repeating the reasonings of
Sect.\ref{sec:2}, one can show that the leading term of AO
expansion of $\W_{11}(\alpha; \tau)$ has the behavior of
$1\!\left/ \alpha^2 \right.$. So, after integration with the
weight function $\varphi(\tau)$ the first term in the square
brackets in (18) makes a contribution of order $O(1)$. The leading
terms in $\W_{12}$ and $\W_{1}^{2}$ both are $O(1\!\left/ \alpha^3
\right.)$. So, the second term in the square brackets and the last
term in (18) are of order $O(\alpha)$. By a similar reasoning one
can show that the neglected terms in (18) are of the order
$O(\alpha^2)$. So in view of $\W_{1}(\alpha; \tau) =
O(\alpha^{-1})$, formula (18) approximates $\W (\alpha; \tau)$
within the NNLO precision.

It should be noted, that the above reasoning is valid only as long
as the functions $\Sigma_{l}(\tau)$, $l=2,3,\cdots$, and several
of their derivatives are regular functions in some neighborhood of
$\tau = 0$. If this is not the case, then the products of
$\Sigma_{l}(\tau)$ on $\W_{1n}^m$ must be considered as new
generalized functions, of which the properties should additionally
be investigated. Such situation occurs when the unstable particle
interacts with massless particles (photons). This difficulty may
be eliminated by the inclusion of the regularizing mass for
massless particles, because then functions $\Sigma_{l}(\tau)$
become regular in a vicinity of $\tau = 0$.

Assuming this trick let us now consider the complete AO expansion
of $\W_{1n}^m$. For brevity we omit the corresponding derivation
since it is similar to the one considered in the previous section.
The desirable accuracy of the expansion is controlled by formula
(18), so that $\W_{11}$ should be expanded up to $O(1)$
corrections, and $\W_{12}$ and $\W_{1}^{2}$ up to $O(\alpha^{-2})$
corrections. However, in further calculations major precision is
required. So, let us consider
\begin{equation}\label{19}
{\W}_{11}(\alpha ;\tau )= E(\tau )+{1 \over {\tau ^3}}+O(\alpha
),\;\;\,{\W}_{12}(\alpha ;\tau )= E(\tau
)+O(1),\;\;\,{\W}_1^2(\alpha ;\tau )= E(\tau )+O(1) .
\end{equation}
Here in all three cases $E(\tau)$ is defined by (9), but with
different coefficients $c_n$. In the compact form, when $c_n$ are
defined as functions on $\tau$, we get:
\begin{tabbing}\label{20,21,22}
mmmmmmm \= mmmmmmmmmm \= mmmmmmmmmmm \= mmmmmmmmmmmmmmmm \= (20)
  \kill
  ${\W}_{11}:$ \>
  $c_0=\frac{\displaystyle \i \pi }{\displaystyle 2\,\alpha
  ^2f^2}\,,$ \>
  $c_1=\frac{\displaystyle \pi ( {\i h+f} )}
  {\displaystyle 2\,\alpha f^2}\,,$ \>
  $c_2=\frac{\displaystyle \pi ( {\i h^2+\i f^2+2hf} )}
  {\displaystyle 2f^2}\,;$ \>
  (20) \\[0.5\baselineskip]
  ${\W}_{12}:$ \>
  $c_0=-\frac{\displaystyle \pi}{\displaystyle 4\,\alpha ^3f^3}\,,$
  \>
  $c_1=\frac{\displaystyle \pi ( {\i f-h} )}
  {\displaystyle 4\alpha ^2f^3}\,,$ \>
  $c_2=\frac{\displaystyle \pi ( {2\,\i h f+f^2-h^2})}
  {\displaystyle 4\,\alpha \kern 1pt f^3}\,;$ \>
  (21) \\[0.5\baselineskip]
  ${\W}_1^2:$ \>
  $c_0=\frac{\displaystyle \pi}{\displaystyle 2\,\alpha ^3f^3}\,,$ \>
  $c_1=\frac{\displaystyle \pi h}{\displaystyle 2\,\alpha ^2f^3}\,,$
  \>
  $c_2=\frac{\displaystyle \pi ( {h^2+f^2} )}
  {\displaystyle 2\,\alpha f^3}\,.$ \>
  (22) \\[-0.8\baselineskip]
\end{tabbing}
\setcounter{equation}{22} Notice here that the coefficients
singular in $\alpha$ do not depend on the prescription for the
poles in $\tau$, since in the examples considered above the
non-integrable terms (for which the prescription is needed) are
non-singular in $\alpha$.

Based on the derived results we formulate the following properties
of the expansion. (All of them can be verified by direct complete
expanding and comparing the results.)

\medskip \noindent {\underline \large \it \underline{Property 1.}}
The incomplete expansion (18) is equivalent within the given
precision to the complete AO expansion of $\W (\alpha; \tau)$.

\medskip
The next properties are non-trivial and some additional argument
should be given.

\medskip
\noindent {\underline \large \it \underline{Property 2.}} Within
the given precision an incomplete expansion of $\W (\alpha; \tau)$
is equivalent to its complete AO expansion if this incomplete
expansion includes as Dyson resummed all contribution non-zero at
$\tau = 0$ to $\Im \Sigma_1 (\tau)$. All other contributions to
$\Sigma_1 (\tau)$ may be transferred from denominators to
numerators in the sense of a conventional expansion. Moreover,
only a finite number of terms of the latter expansion are relevant
within the given precision (see remark at the end of the proof).

We perform the proof in two steps. At first we will show that all
contributions to $\Sigma_1 (\tau)$ zero at $\tau = 0$ without loss
of precision may be expanded in the conventional sense. Then we
will show that the same operation may be done also for the whole
of the real part of $\Sigma_1 (\tau)$.

So, let $\Sigma_1 (\tau) = \Sigma_{01} (\tau) + \widetilde\Sigma_1
(\tau)$, where by definition $\widetilde\Sigma_1 (0) = 0$, but
$\Sigma_{01}(0) \not= 0$. Since in some neighborhood of $\tau = 0$
$\widetilde\Sigma_1 (\tau)$ is a correction to $\Sigma_{01}
(\tau)$, its contribution may be expanded like $\Sigma_n (\tau)$
with $n>1$ in (18):
\begin{equation}\label{23}
\matrix{\\[0.5\baselineskip] {\W}(\alpha ;\tau )={\W}_1(\alpha
;\tau )+\left[ \matrix{\left( {\alpha \,\widetilde \Sigma
_1+\alpha ^2\Sigma _2+\alpha ^3\Sigma _3}
\right)\,{\W}_{11}(\alpha ;\tau )\hfill\cr +\left( {\alpha
^2\widetilde \Sigma _1^2+2\alpha ^3\widetilde \Sigma _1\Sigma
_2+\alpha ^4\Sigma _2^2} \right)\,{\W}_{12}(\alpha ;\tau) +
\mbox{h.c.}\hfill\cr} \right]\,\hfill\cr \qquad +\left[ {\alpha
^2\left| {\widetilde \Sigma _1} \right|^2+\alpha ^3\left(
{{\stackrel{\sim}{\Sigma}}_1 \stackrel{\ast}{\Sigma}_2 +
\mbox{h.c.}} \right) + \alpha^4\left| {\kern 1pt \Sigma _2}
\right|^2} \right]\,{\W}_1^2(\alpha ;\tau )+O(\alpha
^2)\;.\hfill\cr}
\end{equation}
Here symbol ``$\ast$'' means complex conjugation. In the
denominators of $\W_{1n}^m$ only the $\Sigma_{01} (\tau)$ is Dyson
resummed. The remainder in (23) is estimated in the AO sense. The
AO expansions of $\W_1$, $\W_{11}$, $\W_{12}$ and $\W_{1}^{2}$
within the required precision were previously described in this
section.

The proof of (23) may be given noticing that within the given
precision the quantity $\alpha\widetilde\Sigma_1$ gives a non-zero
contribution only if it is raised to not more than the second
power. In fact, against the background of the regular terms this
property is obvious. If $\widetilde\Sigma_1^2$ (or
$|\widetilde\Sigma_1|^2$) is multiplied by a counterterm
originating from $\W_{1n}^m$, the result is non-zero only in the
presence of second- or higher-order derivatives of the
delta-function in the counterterm (otherwise there acts the
property $\widetilde\Sigma_1 (0) = 0$). Consider the functions
$\W_{1n}^m = \left[\W_1\right]^m \! \Delta_1^n$, $n \ge 0$, $m \ge
0$, $n+m-1$ is the number of self-energy insertions on one side of
the cut of the diagram of unitarity, $m-1$ is the number on the
other side. In these functions such counterterms appear in order
$\alpha^{-(n+2m-1)} \times \alpha^2$, and only those functions
$\W_{1n}^m$ are to be multiplied by $\widetilde\Sigma_1^2$ (or
$|\widetilde\Sigma_1|^2$) which satisfy the condition $n+2m-2 \ge
2$. Note that in both insertions of $\widetilde\Sigma_1$, each
gives a factor $\alpha$. Of the other remaining $n+2m-4$
insertions of $\widetilde\Sigma_k$, $k \ge 2$, each gives a factor
not less than $\alpha^2$. Thus all considered terms result in
$O(\alpha)$ contribution. While extending the above reasoning to
the third and the higher powers of $\alpha\widetilde\Sigma_1$, we
easily see that they give nonzero contributions of $O(\alpha^2)$
only.

The above result may be generalized to the real part of
$\widetilde\Sigma_1 (\tau)$. In this case $\widetilde\Sigma_1
(\tau)$ in (23) is to be defined by $\Im \widetilde\Sigma_1 (0) =
0$ with $\Im \Sigma_{01} (0) \not= 0$. The basis for this
generalization is the observation that the real part of the
self-energy does not contribute to the leading-order term of the
AO expansion of $\W (\alpha; \tau)$. Let us remark, that at first
sight formula (23) with this modification should look much mo\-re
complicated, because from a formal point of view it should contain
an infinite series of terms of $[\alpha\,\Re \Sigma_1 (\tau)]^n
\times \W_{1n}$, each of which is of order $O(\alpha^{-1})$, etc.
However, forming the groups with other functions $\W_{1n}^m$ all
superfluous terms must mutually cancel. The mentioned groups will
be formed by virtue of relations of the type $2\,\Re\W_{12} +
\W_1^2 = O(\alpha^{-1})$ [not $O(\alpha^{-3})$], etc.

It is worth noticing that the above result permits one to expand
the bosonic corrections to self-energy of W- and Z-bosons within
the \emph{finite} number of terms. (This is because the mentioned
bosonic corrections possess the property
$\Im\Sigma_{1}^{\mbox{\scriptsize bos}}(0) = 0$
\cite{B-P,Bardin}.) This is a nontrivial result since according to
the formal power counting all these terms are of the same order in
$\alpha$ and, consequently, should be explicitly taken into
consideration.

\medskip
\noindent {\underline \large \it \underline{Property 3.}} There is
the following approximation of $\W (\alpha; \tau)$ up to
$O(\alpha^n)$ corrections:
\begin{equation}\label{24}
{\W}(\alpha ;\tau )={\W}_{[n]}(\alpha ;\tau )-\alpha ^{n-1}{{\Im
\Sigma _{n+1}(0)} \over {\left[ {\Im \Sigma _1(0)}
\right]^2}}\>\pi \delta (\tau )+O(\alpha ^n) .
\end{equation}
Here $\W_{[n]}(\alpha; \tau)$ stands for the propagator squared
with Dyson resummed contributions to the self-energy up to
$n$-loops. The second term includes the $(n+1)$-loop correction
which is necessary (with the appropriate factors) within the
indicated precision. Formula (24) follows immediately from the
obvious generalization of (18) to the case of any $n$ taking into
consideration the first result in (20).

It is worth noticing that the second term in (24) makes a
contribution in the highest order within the given precision.
Consequently, any $O(\alpha)$-variation of the mass is worthless
in this term, because the effect may be referred to the neglected
term of order $O(\alpha ^n)$. Actually this is a very important
observation, since in the case of electroweak theory it permits
one to leave out of account the ambiguities in the definition of
the gauge-bosons masses \cite{Sirlin}. In particular, one may
consider the factor $\Im \Sigma_{n+1}(0)$ to be defined at the
complex-value pole position in the spirit of the pole-scheme (as a
result $\Im \Sigma_{n+1}(0)$ becomes completely gauge-invariant),
or simply as the $n$-loop correction to the width of the unstable
particle divided by its mass.

Formula (24) represents the quantitative characteristic of the
property of the loss of one-loop PT order in the resonance region.
Indeed, as long as $\W_{[n]}(\alpha; \tau) \sim \alpha^{-1}$ at
$\alpha \to 0$, the second term in (24), represents the $n$-th
order correction, but not the $(n+1)$-th one, as might naively be
expected. Since the second term in (24) cannot be obtained from
the analysis of an amplitude, it is pertinent to call it the
\emph{anomalous additive term}.

\medskip
\noindent {\underline \large \it \underline{Property 4.}} The
above expansions possess the following transformation property
when the argument $\tau$ is shifted by a quantity of order
$O(\alpha)$:
\begin{equation}\label{25}
\widetilde {{\W}}(\alpha ;\tau )=\widetilde {{\W}}(\alpha ;\tau
-\alpha \,m^2)_{\left| {\scriptstyle {h(\tau - \alpha \,m^2)\to
h(\tau )-m^2}\hfill\atop \scriptstyle {f(\tau -\alpha \,m^2)\to
f(\tau )}\hfill} \right.} .
\end{equation}
Here $\widetilde{\W}(\alpha; \tau)$ stands for any incomplete or
complete AO expansion of the considered above functions; the
quantity $m^2$ is of order $O(1)$.

Property (25) means the non-sensitivity of the entire formalism
with respect to variation of the mass shell within $O(\alpha)$.
Moreover, it means independence of the formalism from the UV
renormalization scheme. Indeed, at the one-loop level the
transition, for example, from the MS scheme to the on-mass-shell
(OMS) scheme is described by \footnote{ Remember, we do not
consider contributions to the numerators of propagators.
Therefore, we disregard the multiplicative wave function
renormalization.}
\begin{equation}\label{26}
\matrix{\\[0.5\baselineskip] \Sigma _{OMS}(p^2)=\Sigma (p^2)-\Re
\Sigma (M_{OMS}^2)-(p^2- M_{OMS}^2)\times \Re
\Sigma'(M_{OMS}^2)\;,\hfill \cr M_{OMS}^2=M^2-\Re \Sigma
(M_{OMS}^2)\;.}
\end{equation}
It is obvious that (26) belongs to the class of transformations
covered by (25). The transformation at the multi-loop level is
controlled by formula (18). So this can be done in accordance with
the standard recipes of the UV renormalization, which do not
depend on presence of the ``infrared'' counterterms located on the
mass shell (see also \cite{PNP} and references therein). Let us
note that the transformation to another scheme of UV
renormalization may be done (speculatively) before the squaring of
the amplitude and the AO expansion. The consequent AO expansion by
no means ``feels'' in what scheme the Green functions have been
renormalized.

The property (25) is trivial in cases of non-expanded $\widetilde
{\W}$ or its formal expansions. The nontrivial aspect is that it
remains valid for the counterterms $E(\tau)$, too. However, this
also can be understood if one notes that the transformation $\tau
\to \tau - \alpha m^2$ does not affect the structure of the
homogenization at the scaling $\tau \to \xi\tau$, $\alpha \to
\xi\alpha$ (see Sect.\ref{sec:3}). As a result, the property (25)
is valid in the most general case.

\smallskip

Finally, we present one more property which represents doubtless
independent interest.

\medskip
\noindent {\underline \large \it \underline{Property 5.}} From
(25) follows the next recurrent formula for the coefficients
$c_n$,
\begin{equation}\label{27}
c_{n-1}={1 \over n}\left[ {{1 \over \alpha }{\partial  \over
{\partial \kern 1pt h_0}}- \sum\limits_{r=0}^{N-n} {\left(
{h_0^{(r+1)}{\partial  \over {\partial \kern 1pt
h_0^{(r)}}}+f_0^{(r+1)}{\partial  \over {\partial \kern 1pt
f_0^{(r)}}}} \right)+{\partial  \over {\partial \kern 1pt M^2}}}}
\right]\,c_n .
\end{equation}
Here $c_n=c_n \bigl( {M^2;\alpha ;h_0,\ldots
h_0^{(N-n)};f_0,\ldots f_0^{(N-n)}} \bigr)$ are the constants
independent from $\tau$, $n$ runs values $0 \le n \le N$, where
$N$ is the maximum degree of the derivative of the delta-functions
in $E(\tau)$. Formula (27) is written down with possible
dependence of the coefficients $c_n$ on the parameter $M^2$. An
essential point for the derivation of (27) is the expansion in
powers of $\alpha$ of the delta-function $\delta(\tau - \alpha
m^2)$ and its derivatives in r.h.s. of (25).

The practical value of (27) consists in the opportunity to check
the results of calculation of counterterms, or to determine the
``lower'' coefficient $c_{n-1}$ to within $O(\alpha^L)$ if the
``higher''coefficient  $c_n$ is known to within $O(\alpha^{L+1})$.

\section{Massless particles exchange}
\label{sec:5}

The problem of taking into consideration massless particles
(photons) requires a special analysis because massless-particle
contributions to the self-energy of a massive (unstable) particle
involve the singularity of $\tau \! \times \ln (\tau - \i 0)$. The
first derivative of this expression is not defined at zero. So
already the first corrections to coefficients (15) become
uncertain.

One way to solve this problem is to introduce a regularizing mass
for massless particles (a soft-photon mass). Then the
non-analyticity at $\tau = 0$ disappears, and this fact opens a
way to use without problems the above-derived formulas of the AO
expansions. In a properly defined probability, taking into account
radiation of the real soft photons, all contributions singular in
the photon mass should cancel among themselves. That leads to
continuity of probability with respect to the photon mass. So,
while solving qualitative problems, one could not worry about the
presence of the photon mass, because in the final results the
dependence on it may be eliminated by the ordinary passage to the
zero-mass limit.

It should be stressed that in the above discussion the photon mass
should not necessarily be infinitesimal. Moreover, it may be
chosen to be so large as far as is needed for the solution of a
particular qualitative problem. In what follows we use the
photon-mass insertion for the consistent determination of the
orders of expansion in the coupling constant of the photon
contributions to the probability. After completion of all
cancellations and after the passage to the zero-mass limit these
orders will not change. It is worth noticing that the photon mass
may be introduced in such a way as to guarantee validity of WI.
This can be seen by consideration of the problem in the
Stueckelberg formalism.\footnote{ The author is grateful to
A.A.Slavnov for indication of this fact} Moreover, a photon mass
(and a gluon mass, as well) may be introduced in a totally
gauge-invariant fashion \cite{Slavnov}. So, the photon mass will
not brake down gauge cancellations.

Another way \cite{Base} to solve the problem of massless particles
is based on usage of the regularization property of the parameter
$\alpha$. This method is deeply involved in the context of the AO
expansion. It should be noted, however, that it is able to cure
only those IR divergences of which the origin is connected with
the emergence of an additional singularity in $\alpha$ at $\alpha
\to 0$. In reality such divergences appear only in the diagrams
where the soft momenta of massless particles come into the
propagator of unstable particle considered on shell. Let us note
that cancellation of these IR divergences means cancellation of
the corresponding singularities in the coupling constant, and vice
versa. The idea of the method of \cite{Base} consists of a
stepwise expansion of the full Green functions squared: first in
the contributions of the massless particles only, and then in
other vertices using the AO technique. (The first step will not
lead to an infinite series of equal orders in $\alpha$ in the
resonance region owing to Property~2 and the property
$\Im\Sigma_{1}^{\mbox{\scriptsize bos}}(0) = 0$ for the
massless-particle contributions.)

Acting in the framework of the method of \cite{Base}, after the
first step one gets the modified Green functions with unstable
propagators not containing contributions of the massless
particles. However, this simplification is not indisputable, since
there will appear a lot of configurations with the explicit
contributions of photons for which some special counterterms will
be needed. (In fact they will regularize the products $\Sigma_l
\times \W_{1n}^m$; see the discussion in the previous section.)
Basing ourselves on considerations of unitarity one can justify
cancellation of the considering class of IR divergences
\cite{Base}. However, the property of the gauge cancellations
still remain to be proved. So this method of handling the IR
divergences seems not too good, especially since it does not
provide a complete solution of the problem and requires a lot of
additional efforts.

\section{The gauge cancellations}
\label{sec:6}

Now we are ready to give a proof of the gauge cancellations in any
order of the modified PT. We present our argumentation in a rather
general form, applied basically to any unstable particle in
electroweak theory. The main idea is to separate in the
probability the contributions certainly possessing the property of
gauge cancellations from the problematic contributions. That
allows us to concentrate on study the problematic contributions
only.

Let us begin with some preliminary notes. First of all we
determine the photon-mass regularization for IR divergences. The
advantages of this method have been discussed in the previous
section. Here we stress that this method permits one to consider
soft photons like ordinary massive particles. That means that
their contributions either may be referred completely to vertex
blocks, or, in other cases, they suppress the resonant behaviour
of the corresponding contributions. Really, an exchange by a
massive particle suppresses the \emph{on-shell} propagating of
unstable particles both before and after the emission of the
massive particle. As a result, the problem of instability is
reduced solely to configurations which include only the pairs of
unstable propagators of equal mass and momenta, placed on both
sides of the cut of unitarity. It is worth remembering that the
mass insertion for photons will not violate gauge cancellations
(see previous section). Moreover, since the probability is a
continuous function of the photon mass the dependence on it may be
eliminated in the final results by the ordinary passage to the
zero-mass limit.

The next note concerns the unphysical pole contribution to the
vector boson propagators. Assuming the parameterization for the
self-energy
\begin{equation}\label{28}
\Sigma_{\mu \nu }(p)=\Sigma \,g_{\mu \nu }+\Sigma_L\,p_\mu p_\nu ,
\end{equation}
we come to the following explicit expression for the full
propagator in $R_{\xi}$-gauge:
\begin{equation}\label{29}
D_{\mu \nu }(p)={{g_{\mu \nu }-{{p_\mu p_\nu } \mathord{\left/
{\vphantom {{p_\mu p_\nu } {p^2}}} \right.
\kern-\nulldelimiterspace} {p^2}}} \over {p^2\!- M^2+\Sigma
}}+\xi\, {{{{p_\mu p_\nu } \mathord{\left/ {\vphantom {{p_\mu
p_\nu } {p^2}}} \right. \kern-\nulldelimiterspace} {p^2}}} \over
{p^2\!-\xi M^2+\xi \left( {\Sigma\!+\!p^2\Sigma_L} \right)}} .
\end{equation}
Here the first term represents the product of the spin factor on
the scalar propagator $\Delta (\alpha; \tau)$ introduced in (1).
The second term describes the unphysical contribution. In the
conventional PT it is cancelled due to WI by contributions of the
other unphysical states. Consequently, the second term in (29)
does not lead to non-integrable singularity in the phase space. In
the modified PT this fact allows us to take into account the
second term in the conventionally expanded form. Then, the
property of gauge cancellations will be controlled by the
properties of the first term, taken in an absolute value and
having been squared. Note that the cross terms resulting from the
squaring of the propagator (29) will not lead to non-integrable
singularities in the phase space. So they also may be taken into
account as conventionally expanded.

Now let us show that the first term in (29), being taken into
consideration in framework of the modified PT, does not break the
gauge cancellations in the probability. Remember that in
accordance with (24) its contribution may be presented in the form
of a sum of two terms, where one is conventional and the other is
anomalous. Let us start from analysis of the anomalous term
contribution. Due to its additivity the simplest way to perform
the analysis is to assume for a moment that the squared unstable
propagator consists in this anomalous term only.

Firstly we consider the case with a single unstable particle
production. Let us note that the complementary part of diagram of
unitarity with respect to the given propagator squared, by virtue
of the delta-function \emph{without a derivative} in the anomalous
term, is taken strictly on mass shell. Moreover, the details
\cite{Sirlin} of the definition of the mass shell are inessential
here and may be omitted. (This is because these details have
meaning of corrections, but the anomalous term describes the
highest order within the given precision.) From these two facts
follows the conclusion that the sum of the complementary parts is
gauge invariant, since actually it coincides (up to factors) with
the product of two $S$-matrix elements squared, with one
describing the on-shell production of the unstable particle and
another one describing its decay. Furthermore, in the anomalous
term itself one may neglect the gauge-dependent contributions, if
in this case they are in $\Im \Sigma_{n+1}(0)$, since they are
beyond the given precision (see Property~3 and the discussion
therein). So the total sum of unitarity diagrams that include the
anomalous term within the given precision is gauge invariant.

In the case of multiple production of unstable particles let us
single out any unstable particle and, again, consider its
contribution of an anomalous term. Since the anomalous term is of
the highest order, the contributions of all other unstable
particles should be considered in the leading order of the AO
expansion. By virtue of (7), (9), and (15) the leading-order
contributions are determined by the delta-function \emph{without a
derivative} with a coefficient including the one-loop $f_0$ only
as a non-trivial factor. In view of the gauge invariance of the
one-loop $f_0$ (see, for instance, \cite{B-P}) follows, once
again, the gauge invariance of the total sum of complementary
parts of the diagrams of unitarity. So, the gauge invariance in
this case occurs, too. Due to additivity, a similar reasoning may
be repeated for any other unstable particle of multiple
production, and in each case we will reach the property of gauge
invariance of the corresponding contribution to the probability.

Now let us proceed to the analysis of all other contributions,
which do not at all include an anomalous term. Such contributions
are controlled by the first term in (24). Also, all non-resonant
contributions belong to this class of contributions. As a whole,
these contributions are described by the amplitude squared
determined in the conventional approach with Dyson resummation.
The property of gauge cancellations in the amplitude so determined
has already been shown \cite{Background} in the framework of the
background-field formalism. Let us emphasize that the only
condition which has to be observed when considering these
contributions is the Dyson resummation of all corrections to the
self-energies up to $n$-loops together with taking into
consideration all $n$-loop corrections to the vertices. So,
treating $\W_{[n]}(\alpha; \tau)$ in (24) in such a manner, we
automatically get the property of gauge cancellations in the
probability.\footnote{ It should be noted that within the
background-field formalism after all gauge cancellations there may
remain some residual dependence on the quantum gauge parameter.
This phenomenon has been considered in the original work of
\cite{Background}, where it has been stressed also that this
residual dependence will not affect the high-energy behaviour of
the amplitude. An idea for how to solve the problem of the
residual gauge dependence is discussed in Sect.\ref{sec:8}.}

The above discussion completes a construction which provides both
the gauge cancellations and the necessary precision of the
description in the sense of an expansion in the coupling constant.
Basing ourselves on this result one may do the next step:
proceeding to the complete AO expansion of the probability. Then
the property of gauge cancellations within the given precision
must take place as well, because AO only expands in powers of the
coupling constant the expression which possesses the property of
gauge cancellations. With this note we complete the general proof
of gauge cancellations in the approach of the modified PT.

\section{Generalization of the fermion-loop scheme}
\label{sec:7}

If the analysis may be restricted by NLO precision, for example
for the case of W-pair production studied at LEP2, the gauge
cancellations may be proved within the usual formalism, without
applying the background-field method. The key point is the
well-known result on gauge cancellations in the so-called
fermion-loop scheme \cite{Argyres,Fermion-loop}. Recall that it
consists of including all fermionic one-loop corrections in
tree-level amplitudes and Dyson resumming self-energies. The
difficulty of this scheme is twofold. First, it is not known how
to incorporate the one-loop bosonic corrections into this scheme
without spoiling the gauge invariance. Second, there is a problem
with gauge invariance while taking into account the two-loop
corrections to self-energies in the denominators of unstable
propagators, which is necessary in the resonance region.

Both these problems may be solved within the modified PT approach
with usage of the AO technique. Let us begin with the problem of
the two-loop corrections. As we have seen above, they can be taken
into consideration without breaking gauge invariance by adding the
anomalous terms to the probability. In view of (24) that may be
done by means of the formula
\begin{equation}\label{30}
{\W}(\alpha ;\tau )={\W}_{[1]}(\alpha ;\tau )-{{\Im \Sigma _2(0)}
\over {\left[ {\Im \Sigma _1(0)} \right]^2}}\;\pi \delta (\tau
)+O(\alpha) .
\end{equation}
Here ${\W}_{[1]}(\alpha ;\tau )$ represents the unstable
propagator squar\-ed with Dyson resummed all one-loop corrections.
Remember that ${\W}_{[1]}(\alpha ;\tau )=O(\alpha^{-1})$ at
$\alpha \to 0$. Therefore, the second term in (30) describes the
highest-order (NLO) correction. So, any ambiguities in its
definition, in particular those which concern the problem of the
gauge invariance, may be referred to the discarded terms of order
$O(\alpha)$. (See Property~3 and the discussion therein.)

The problem of one-loop bosonic corrections may be solved, too.
Let us group these corrections into two classes. To the first
class we refer the corrections to self-energy of unstable
particles. To the second class we refer the corrections to the
vertex factors, the corrections due to exchange processes, and due
to the real (soft) photons.

The corrections of the first class can easily be taken into
consideration by means of the fact that the imaginary parts of the
on-shell bosonic corrections to the self-energies of W- and
Z-bosons are zero \cite{B-P,Bardin}. Owing to this fact and
Property~2 of Sect.~\ref{sec:4} these corrections may be
transferred from the denominators of unstable propagators to their
numerators without loss of precision. Moreover, in an OMS-like
scheme of UV renormalization, where the renormalized self-energies
satisfy the condition $\Re \Sigma _1(0)=\Re \Sigma '_1(0)=0$, one
has the following approximation:
\begin{equation}\label{31}
{\W}_{[1]}(\alpha ;\tau )={\W}_{1F}(\alpha ;\tau )+O(\alpha).
\end{equation}
Here ${\W}_{1F}(\alpha ;\tau )$ represents the propagator squared
where only the fermionic one-loop corrections are Dyson resummed.
Substituting (31) into (30) we come to the formula which leads to
gauge cancellations. In order to see this, we should only repeat
the reasoning of the previous section keeping in mind the gauge
invariance in the fermion-loop scheme.

However, the bosonic corrections of the second class have not yet
been taken into account. In order to do that let us make use the
fact that within the NLO approximation and in the presence of the
\emph{corrections}, the quantity ${\W}_{1F}(\alpha ;\tau )$ must
be taken into consideration in the leading-order approximation
only. Remember that in this approximation $\W_{1F}(\alpha ;\tau
)=\pi\!\left/ (\alpha f_{0F}) \right. \!\times\delta(\tau )$, and
this expression is explicitly gauge invariant. Moreover, since the
delta-function enters without derivatives, the sum of all factors
appearing in its presence in the unitarity diagrams are also gauge
invariant (see previous section).

Thus, we come to the following recipe of the generalization. We
formulate it having in mind the total cross-section for the
typical LEP2 processes CC10, CC11 and CC20, which have been
studied in the framework of the fermion-loop scheme
\cite{Fermion-loop}. In fact, the generalization consists of
adding to the probability two types of corrections.

The corrections of the first type describe the anomalous
contributions. In the NLO approximation they look like the
cross-section of the pair on-shell production of unstable
particles taken in the leading-order approximation, times the
leading-order decay blocks of unstable particles, and times the
``anomalous'' factor. Indeed, the presence of the anomalous factor
means that all other factors should be taken in the leading order.
However, in the leading-order approximation only the
double-resonant subprocesses contribute to $e^{+}e^{-} \to 4f$.
(In fact these subprocesses are of CC03 class.) That follows from
the fact that only these subprocesses include the factor
$1\!\left/ \alpha^2 \right.$ originating from the product of two
unstable propagators squared. Now, let us remember that the
additive anomalous term in (30) includes the delta-function as
well as ${\W}_{1F}(\alpha ;\tau )$ does in the leading-order
approximation. So, the anomalous terms always contribute
``on-shell''.

\begin{figure*}
\hbox{ \hspace*{0pt}
       \epsfxsize=100pt \epsfbox{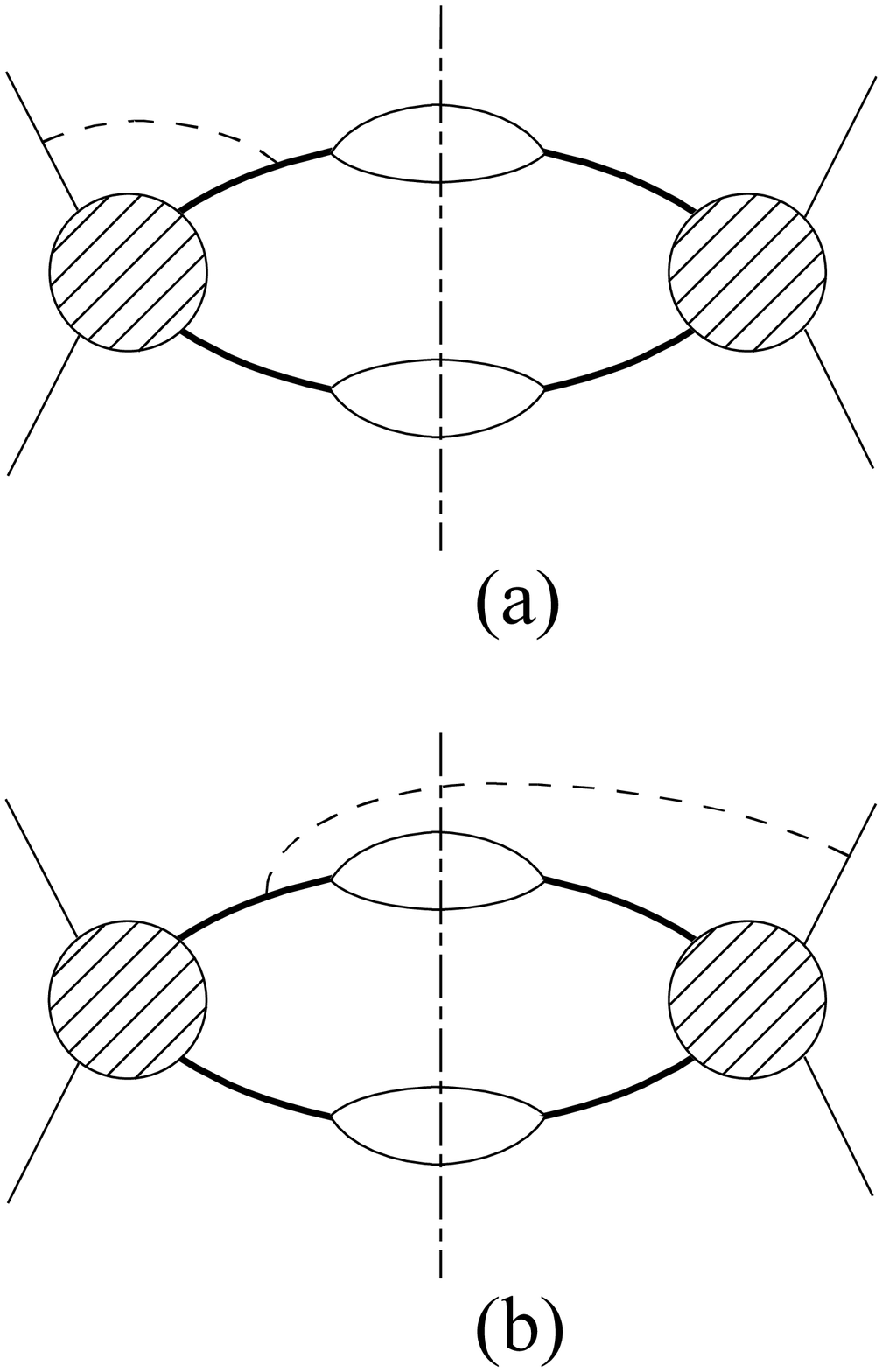} \hspace*{10pt}
       \epsfxsize=100pt \epsfbox{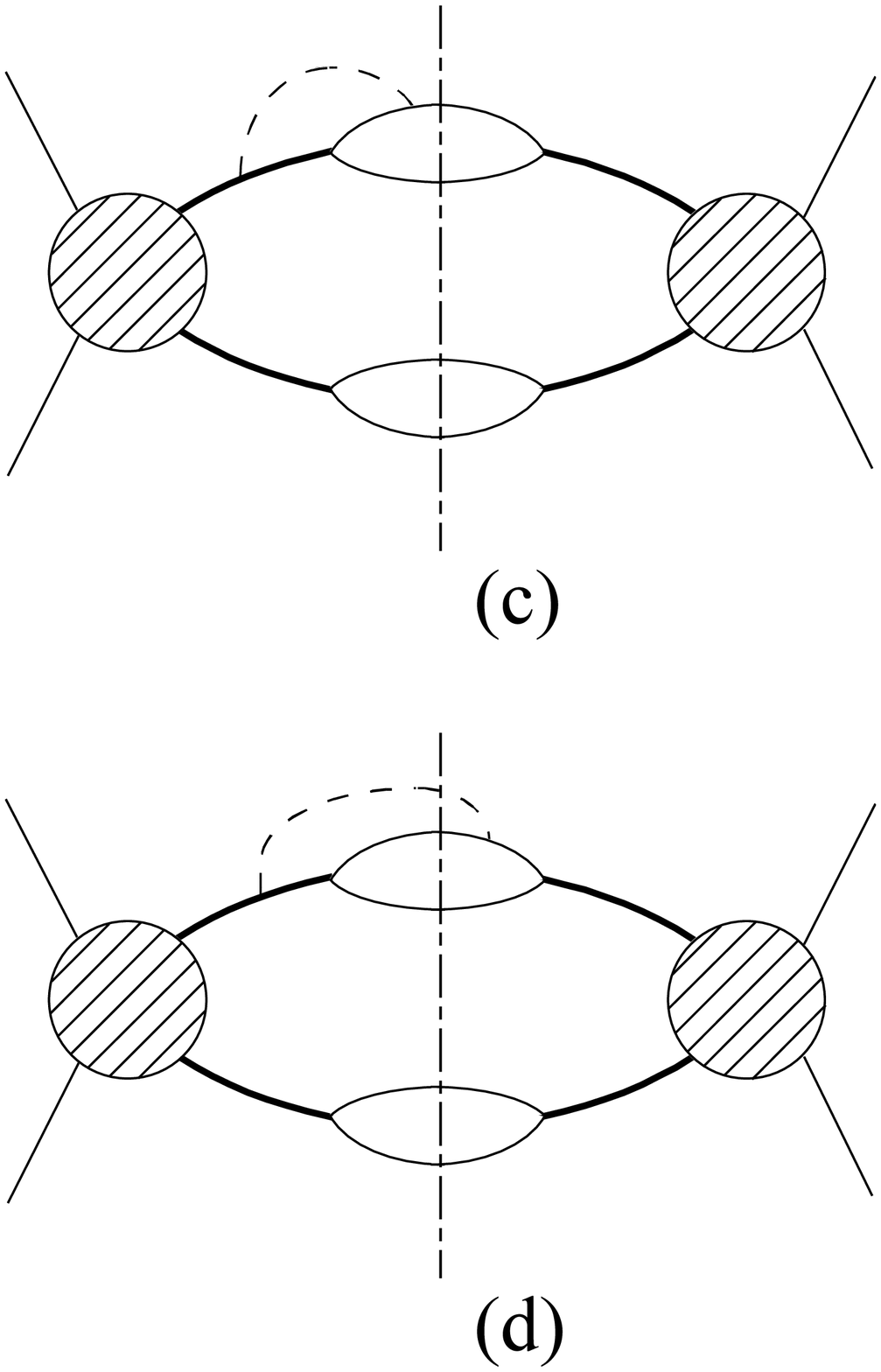} \hspace*{10pt}
       \epsfxsize=100pt \epsfbox{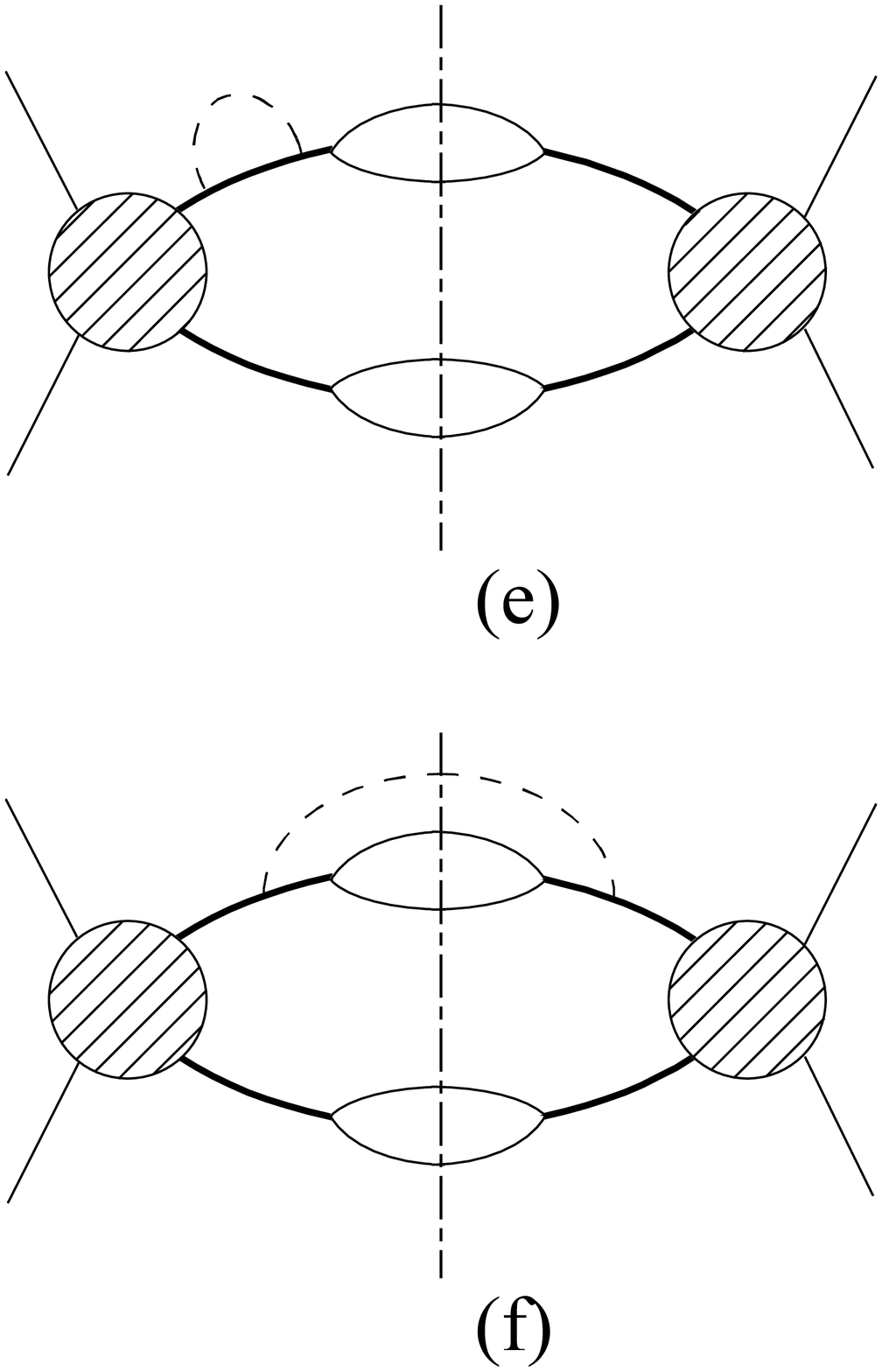} \hspace*{10pt}
       \epsfxsize=100pt \epsfbox{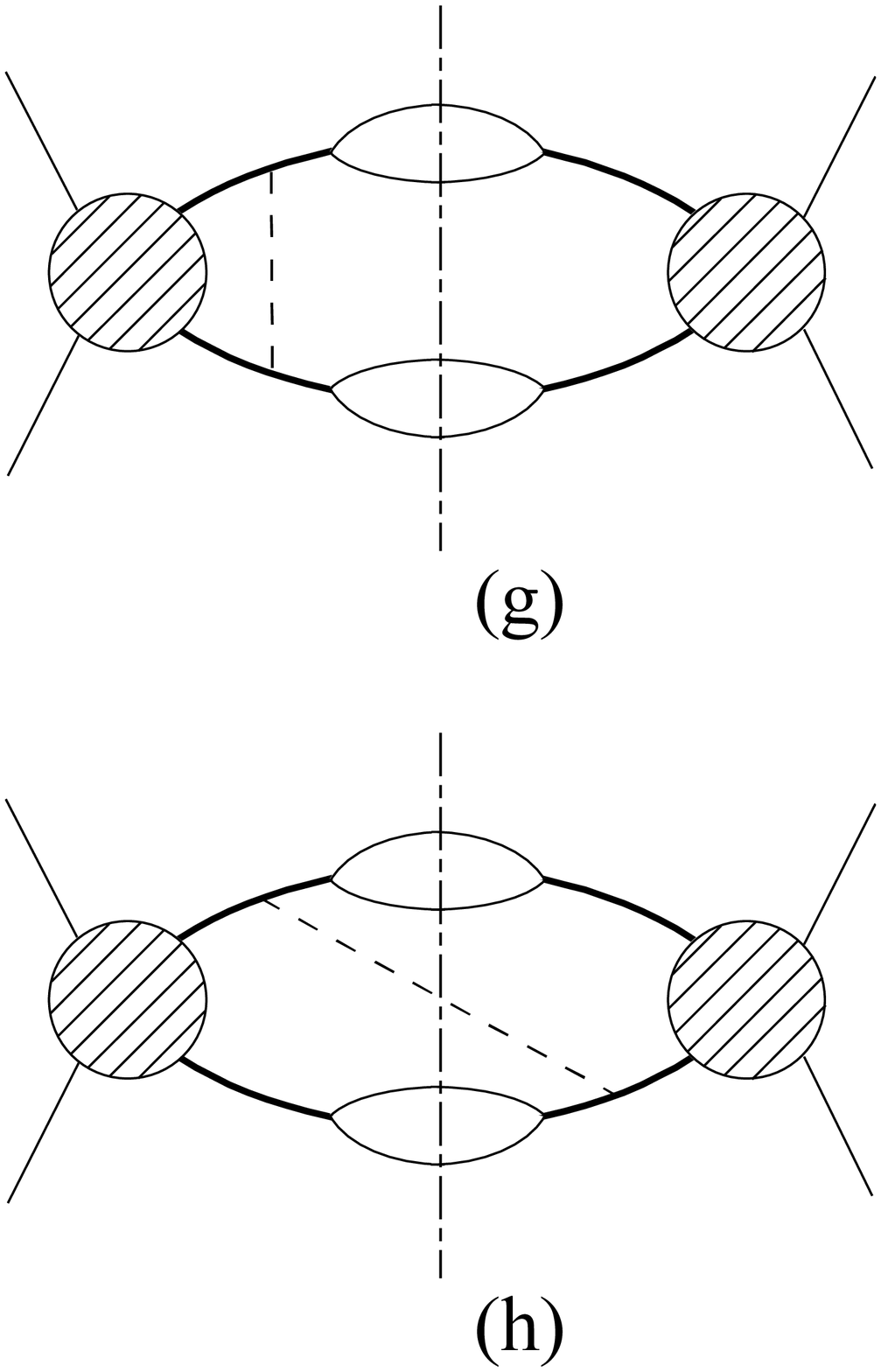} }
\caption{Examples for factorizable and (quasi) non-factorizable
corrections to W-pair production in $e^{+}e^{-} \to 4f(\gamma)$
which have to be taken into consideration in NLO approximation.
(The dashed lines denote massive bosons or soft-photons. The
continuous thin and thick lines represent the initial/final
fermions and the unstable W-bosons, respectively. The vertical
dot-dashed lines indicate the cut of the diagrams of unitarity.
The hatched areas denote the lowest-order Green functions for the
production of the virtual $W$-boson pair.)} \label{fig:1}
\end{figure*}

The corrections of the second type may be represented, again, as
the cross-section of the pair on-shell production of unstable
particles, multiplied by their decay blocks. However, instead of
``anomalous'' factors, they include the bosonic corrections to the
vertex blocks and the corrections due to the real soft-photons.
The examples of the unitarity diagrams which generate
contributions of this type are shown in Fig.~\ref{fig:1}. With the
non-zero mass of the photons all these diagrams include exactly
\emph{two} pairs of unstable propagators of identical mass and
momenta in both sides of the cut. (The only exception,
configuration (1.e), is discussed below.) Therefore, they include
\emph{two} delta-functions of the corresponding kinematic
variables, which make these configurations on shell. The sum of
all such configurations represents the product of the
cross-section of the pair on-shell production of unstable
particles and their decay blocks. Since these quantities are
continuous functions of the photon-mass, the dependence on it may
be eliminated in the very end of the calculation by the ordinary
limiting procedure. Notice that the mentioned property of
continuity follows directly from the well-known theorem for
cross-sections with the real soft-photon contributions.

\begin{figure*}
\hbox{ \hspace*{40pt}
       \epsfxsize=100pt \epsfbox{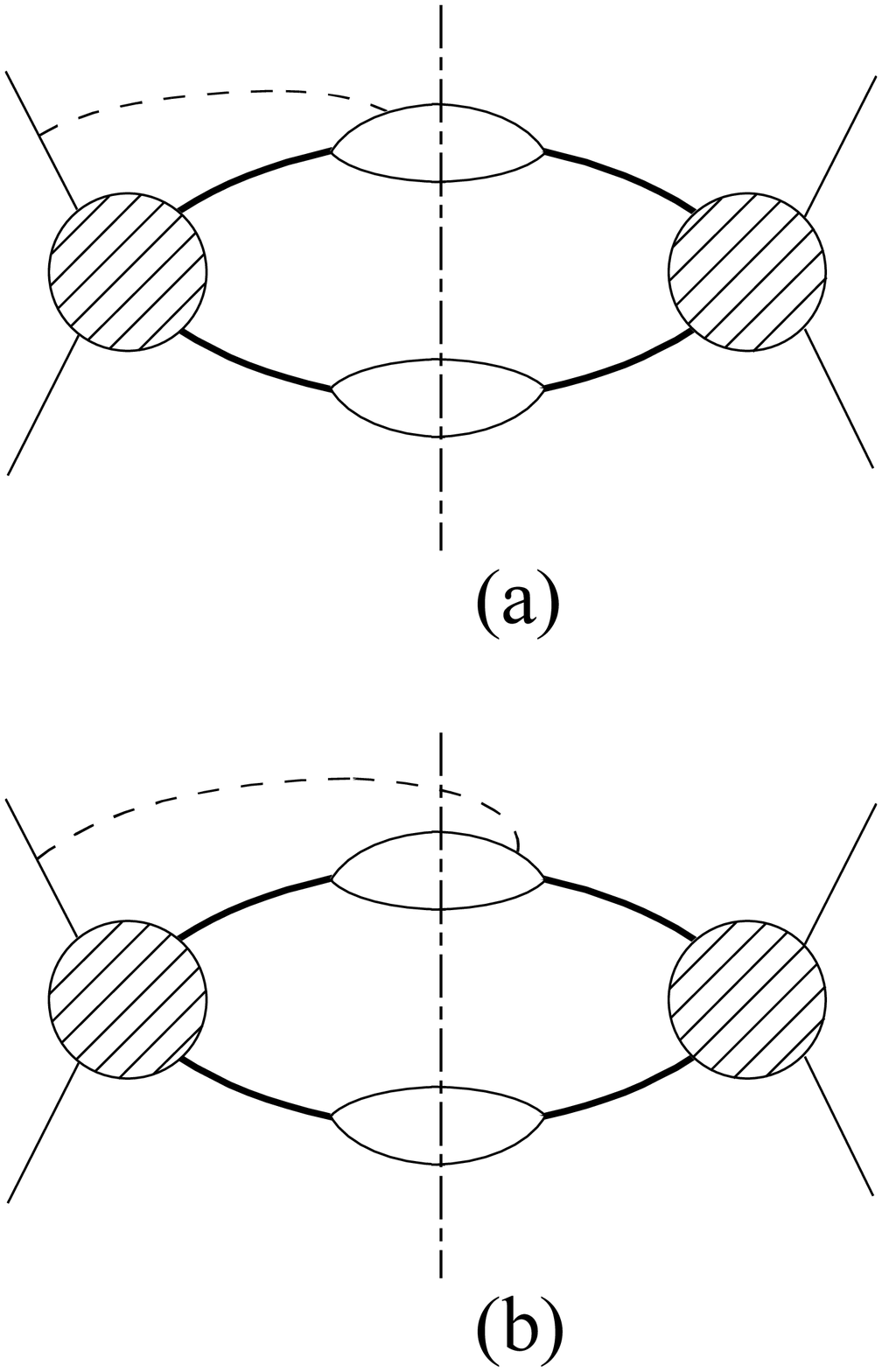} \hspace*{30pt}
       \epsfxsize=100pt \epsfbox{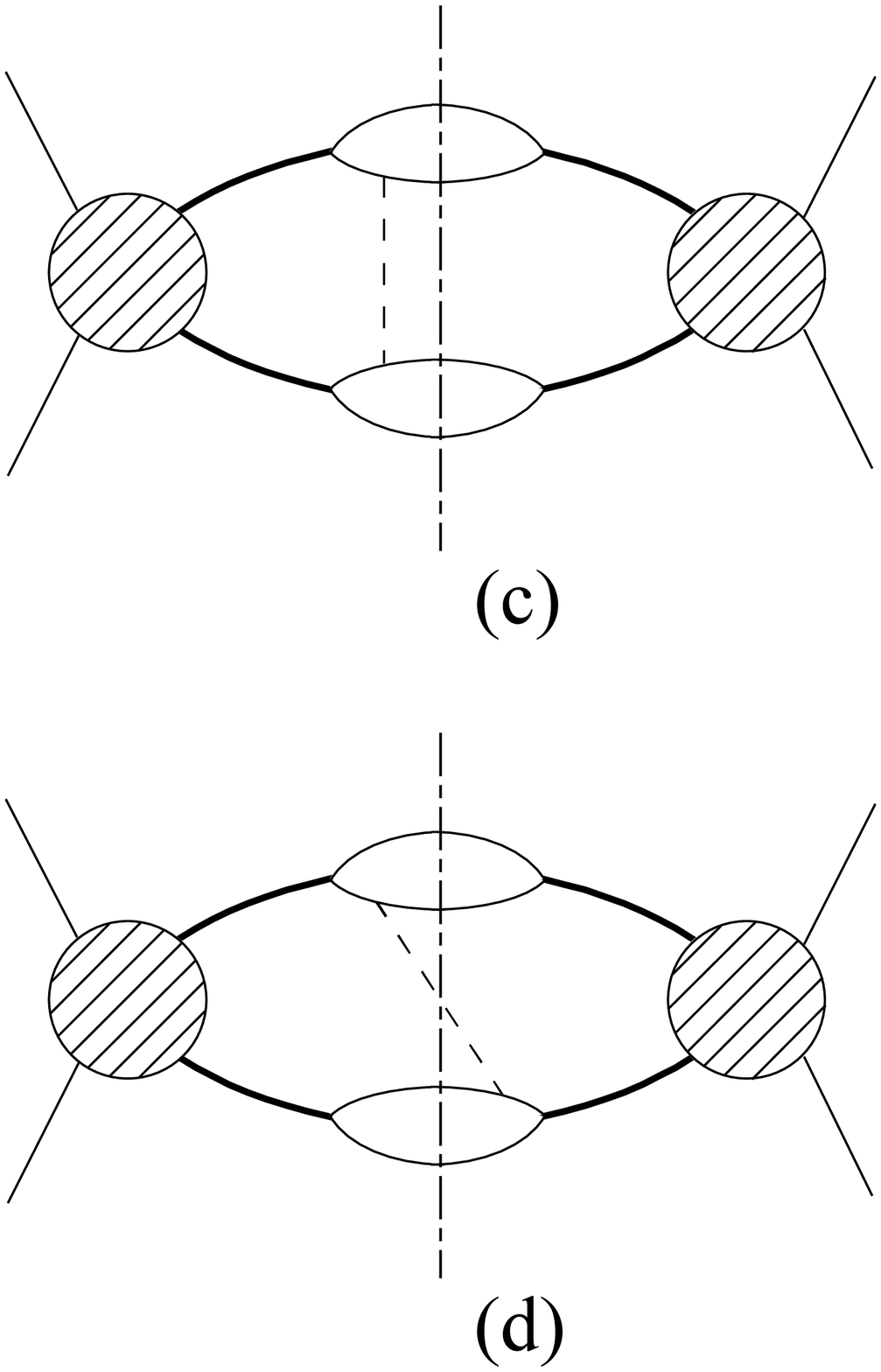} \hspace*{30pt}
       \epsfxsize=100pt \epsfbox{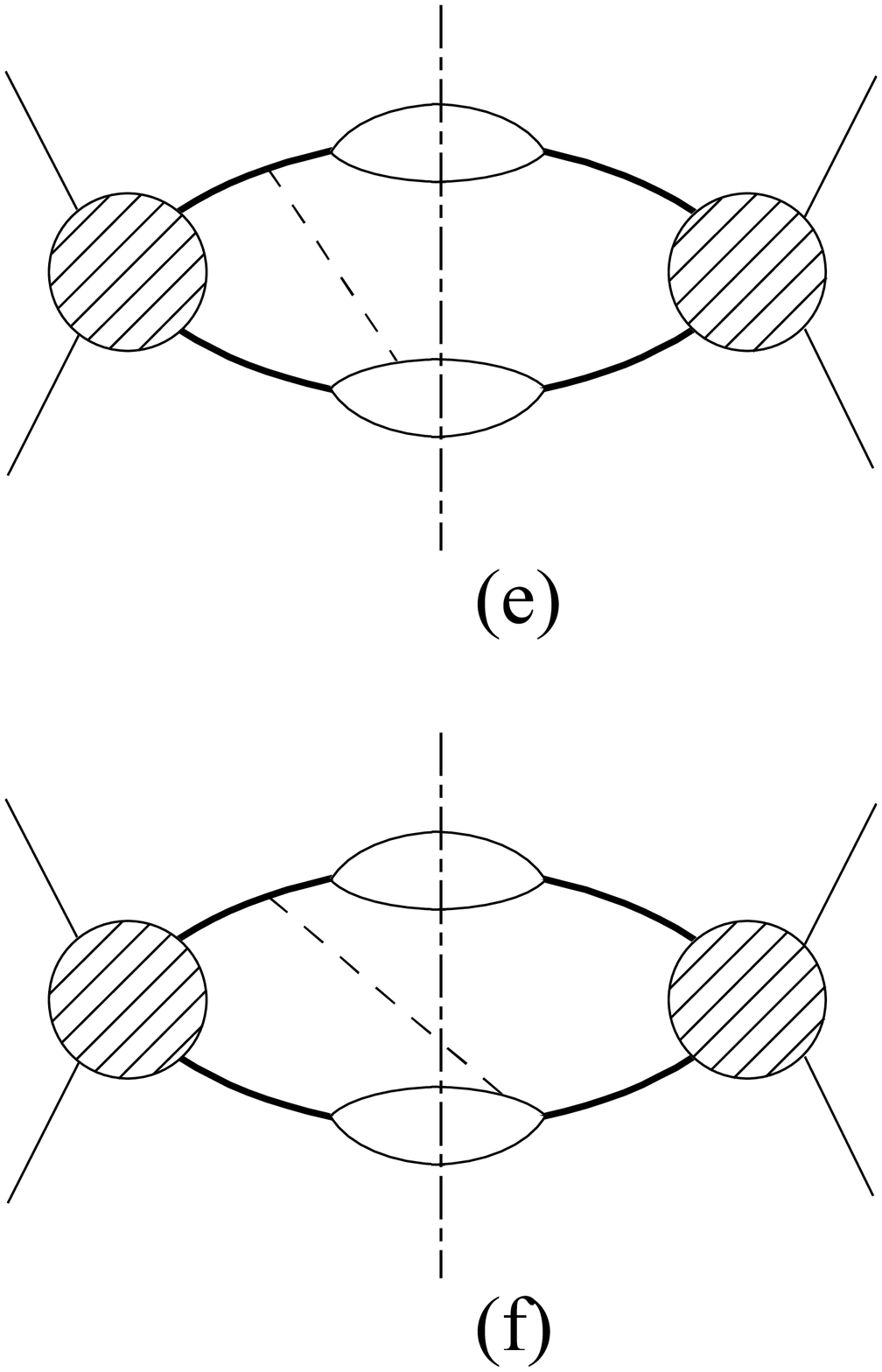} }
\caption{Examples for non-factorizable corrections to W-pair
production which contribute beyond the NLO approximation. }
\label{fig:2}
\end{figure*}

Now let us discuss the mentioned above exceptional configuration
(1.e). Strictly speaking it should not be considered to belong to
the unitarity diagrams, since it describes the self-energy
correction to the unstable propagator, which has already been
taken into consideration by formula (31). Nevertheless, while
considering the cross-section of the pair on-shell production, one
has to take into account the virtual soft-photon insertion to the
external legs, which is due to the wave function renormalization.
The configuration (1.e) was added to the list of diagrams of
Fig.~\ref{fig:1} only in order to indicate this fact.

It should be noted that among the diagrams of Fig.~\ref{fig:1}
there are both factorizable and non-factorizable configurations in
the sense of the classification of
\cite{Doubl-Pole,Dittmaier,Acta}. Nevertheless, at the
intermediate step, when the soft-photons are supplied with the
mass, the non-factorizable configurations of Fig.~\ref{fig:1}
become factorizable. All other non-factori\-zable corrections,
which do not provide this property, produce such configurations
that include \emph{less} than two pairs of unstable propagators of
identical mass and momenta in both sides of the cut. Therefore,
they do not include the leading-order factor $1\!\left/ \alpha^2
\right.$. In view of the presence of the additional factor
$\alpha$ due to the bosonic corrections they contribute beyond the
NLO approximation. The examples of configurations of this type are
shown in Fig.~\ref{fig:2}. It is worth noticing that they may not
be considered as corrections to the cross-section discussed above
of the pair on-shell production of unstable particles, or to their
decay blocks.

The above discussion leads us to the following formulas for the
total cross-section:
\begin{equation}\label{32}
\sigma (s) = \int\limits_{0}^1 \d z \> \phi (z;s) \> \hat\sigma
(zs) ,
\end{equation}
\begin{equation}\label{33}
\hat\sigma (s) = \int\limits_0^s{\d s_+ \!\!\int\limits_0^{(\sqrt
s\;- \sqrt {s_+})^2} {\!\!\d s_- \>\hat\sigma_0(s\,;s_+,s_-)}} .
\end{equation}
Here $\sigma(s)$ is the experimentally measured cross-section at
the center-of-mass energy squared $s$. $\phi (z;s)$ is the
``flux'' function\footnote{In general case, $\phi (z;s)$ includes
experimental cuts and hardware factors.} describing the
contributions of the initial- and final-state photon radiations
with large IR and collinear logarithms, $\hat\sigma(s)$ stands for
the hard scattering improved Born cross-section at the reduced
center-of-mass energy squared (see \cite{LEP2,NuovoCim} and
references therein). It should be emphasized that $\hat\sigma(s)$
contributes like a distribution to $\sigma(s)$, since
$\hat\sigma(s)$ is smeared by the flux function over a wide range
of the kinematic variable ($\phi (z;s)$ is peaked at $z=1$ and has
a tail till a cut at lower values of $z$). Expression (33)
represents $\hat\sigma(s)$ in a form with explicit integration
over the virtualities of unstable particles (over the invariant
masses of the corresponding final states). The quantity
$\hat\sigma_0(s)$ reads
\begin{eqnarray}\label{34}
&\hat\sigma_0(s\,;s_+,s_-) = \hat\sigma_0^{\mbox{\scriptsize
fermion-loop-scheme}}(s\,;s_+,s_-)& \nonumber\\[0.6\baselineskip]
&- \hat\sigma_0^{\mbox{\scriptsize on-shell,tree}}(s\,;M_+,M_-)
\times 2\alpha \,\Im \Sigma_2(0) / \Im \Sigma_1(0) \times
\prod\limits_{\kappa=\pm} \delta(s_{\kappa} - M_{\kappa}^2) \times
B\!R_{\kappa}^{\;\mbox{\scriptsize tree}}&
\nonumber\\[0.5\baselineskip]
& + \hat\sigma_0^{\mbox{\scriptsize on-shell,tree}}(s\,;M_+,M_-)
\times \prod\limits_{\kappa=\pm} \delta(s_{\kappa} -
M_{\kappa}^2)\times B\!R_{\kappa}^{\;\mbox{\scriptsize
{tree/boson-one-loop+real-photon}}}& \nonumber\\[0.5\baselineskip]
& + \hat\sigma _0^{\mbox{\scriptsize {on-shell, boson-one-loop +
real-photon}}}(s\,;M_+,M_-) \times
\prod\limits_{\kappa=\pm}\delta(s_{\kappa} - M_{\kappa}^2)\times
B\!R_{\kappa}^{\;\mbox{\scriptsize tree}} \;.&
\end{eqnarray}%
Here the first term represents the result of the conventional
fermi\-on-loop scheme. All other terms describe its
generalization. Factor 2 in the second term reflects the presence
of two intermediate unstable particles of equal masses;
$B\!R_{\pm}^{\;\mbox{\scriptsize tree}}$ denotes their on-shell
branching in the tree approximation. In the third term one $B\!R$
is determined with the bosonic corrections to the partial width.
(In fact the third term includes the sum of two subterms, with the
modified $B\!R$ for one of two unstable particle.) In (34) we have
used the relation $\alpha \Im \Sigma_1 (0) = M \Gamma_0 (M)$, with
$\Gamma_0 (M)$ being the total on-shell width at tree level. This
relation follows from unitarity and may be verified by direct
calculation \cite{Bardin}.

Formula (34), in principle, may be further simplified by carrying
out the complete AO expansion in its first term. The key formula
for this simplification, written in an OMS-like scheme of UV
renormalization, is as follows:
\begin{equation}\label{35}
{\W}_1(\alpha ;\tau )=\left[\alpha \;{\Im \Sigma _1(0)}
\right]^{-1}\;\pi \kern 1pt \delta (\tau
)+V\!P{1\over{\tau^2}}+O(\alpha).
\end{equation}
Substituting (35) into (34) we finally get
\begin{eqnarray}\label{36}
&\hat\sigma_0(s\,;s_+,s_-) = V\!P\;\hat\sigma_0^{\mbox{\scriptsize
on/off-shell, tree}}(s\,;M_+,s_-) \!\times\!
\delta(s_{+}\!-\!M_{+}^2) \!\times\! B\!R_{+}^{\;\mbox{\scriptsize
tree}}& \nonumber\\[0.5\baselineskip]
& +V\!P\;\hat\sigma_0^{\mbox{\scriptsize off/on-shell,
tree}}(s\,;s_+,M_-)\!\times\! \delta (s_{-}\!-\!M_{-}^2)
\!\times\! B\!R_{-}^{\;\mbox{\scriptsize tree}}&
\nonumber\\[0.5\baselineskip]
&-\hat\sigma_0^{\mbox{\scriptsize on-shell,tree}}(s\,;M_+,M_-)
\times 2\alpha \,\Im \Sigma_2(0) / \Im \Sigma_1(0) \times
\prod\limits_{\kappa =\pm} \delta (s_{\kappa}-M_{\kappa}^2) \times
B\!R_{\kappa}^{\;\mbox{\scriptsize tree}}& \nonumber\\
& + \hat\sigma_0^{\mbox{\scriptsize on-shell,tree}}(s\,;M_+,M_-)
\times \prod\limits_{\kappa=\pm} \delta(s_{\kappa} - M_{\kappa}^2)
\times B\!R_{\kappa}^ {\;\mbox{\scriptsize
{tree/boson-one-loop+real-photon}}}& \nonumber\\[0.5\baselineskip]
&+\hat\sigma_0^{\mbox{\scriptsize on-shell, tree + one-loop +
real-photon}}(s\,;M_+,M_-) \times \prod\limits_{\kappa=\pm} \delta
(s_\kappa-M_\kappa ^2) \times B\!R_{\kappa }^{\;\mbox{\scriptsize
tree}} \;.&
\end{eqnarray}
Here the first two terms and the leading-order tree contribution
in the last term originate from the fermion-loop-scheme term in
formula (34). Namely, the first two terms accumulate the
contributions originating from CC03 subprocesses and
simultaneously from subprocesses of single vector boson
production. In the case of CC03 subprocesses the symbol $V\!P$
means that the corresponding off-shell unstable propagator squared
is approximated, in accordance with (35), by $V\!P\,1\!\left/
\tau^2 \right.$. Another unstable particle in this case is
considered as real, produced on shell. In the case of single
production the vector (unstable) boson is considered as real, too.
The symbol $V\!P$ in this case is superfluous and has to be
omitted.

It should be noted that in the above discussion the usage of
prescription of the principal value in the case of the CC03
subprocesses is necessary, because the usage of any other
prescription for $1\!\left/ \tau^2 \right.$ in (31) and (35) may
change the results in (36). This remark, however, does not concern
the anomalous term, since it arises from the contribution singular
in $\alpha$ to $\W_{11} (\alpha; \tau)$ (see (20) and the note
after (22)).

The above results may be easily generalized to other processes
with unstable particles production, including the processes of the
``NC''-type and of the ``mixed'' type. In the latter case there
appear the additional terms in (34) and (36) of almost the same
structure, that describe the ZZ-contributions (of course,
configurations of Fig.~\ref{fig:1} are not relevant in the
ZZ-case). The generalization for the case of the differential
cross-sections should not lead to difficulties either. However,
the discussion of this question is beyond the scope of the present
paper.

\section{Discussion}
\label{sec:8}

In this paper we have found a practical method that ensures both
the gauge cancellations and a fixed precision of description of
processes mediated by the production and decay of unstable
particles in electroweak theory. The solution has been found in
the framework of the modified PT which consists of an expansion in
the coupling constant directly of probabilities instead of
amplitudes \cite{Base}. In the general case and within any fixed
precision the proposed method is based on the results obtained
earlier \cite{Background} in the framework of the background-field
formalism. Within NLO precision and in the case of W- and/or
Z-pair production we have found a way also in the usual formalism,
basing ourselves on results of the fermion-loop scheme
\cite{Fermion-loop}. But in contrast to the fermion-loop scheme we
perform the description explicit taking into account the bosonic
corrections and the two-loop corrections to the self-energy of
unstable particles in the resonance region. From the practical
point of view the latter result apparently is the most important
one obtained in this paper.

It should be noted that the result on gauge cancellations in the
completely expanded probability generally was expected since the
appearance of \cite{Base}. This work demonstrated that calculation
of the probability of a process mediated by unstable particle
production may be reduced to the regular fixed-order calculation.
The latter fact gave a reason to think that the problem of the
gauge invariance was solved, too \cite{Base}. However, on closer
examination of the problem a large distance to obtaining the
result was found. Indeed, the Feynman rules in the modified PT and
in the conventional PT coincide only out of the mass shell. So,
the gauge invariance a priori occurs only in this area of
kinematic variable, but not in a neighborhood of the mass shell,
where one has the non-standard contributions of the modified PT.
However, due to the delta-functions these non-standard
contributions do not vanish irrespectively of that as far as the
neighborhood is small. Moreover, the properties of these
non-standard contributions are unknown in advance. In particular,
it is not known whether they are gauge invariant or not. This
becomes especially clear in the higher orders of the AO expansion,
beginning with NLO, where the non-standard contributions include
the derivatives of the delta-functions.

In the present paper we avoid this difficulty by making use of
incomplete AO expansions of the unstable propagators squared. This
method allows us to carry out a comparison of the results of the
modified PT with those of the conventional PT with Dyson
resummation. The special role in this scheme is assigned to the
anomalous additive term that corrects the result of incomplete
expansions  of the propagator squared in the vicinity of mass
shell.

We draw attention to a particular problem that arises when
comparing the results obtained in the modified PT with the results
obtained in the conventional approach with use of the
background-field formalism. (It should be emphasized that there is
no such problem in the conventional formalism on using the
fermion-loop scheme.) This is the problem of the residual
dependence on the quantum gauge parameter which still remains in
the background-field formalism after all gauge cancellations
\cite{Background}. In the general case this dependence may pass on
to the results of the modified PT. Nevertheless, \emph{in the
probability} owing to the phenomenon of changing of the order of
particular contributions, which in turn is a corollary of the
singularity in the coupling constant in the propagator squared,
one may expect that this dependence will drop out within the given
precision. Note that we indicate only an opportunity of giving a
solution to the problem, but not the actual solution. However,
earlier it was not clear even how to initiate a solution
\cite{Fermion-loop,Background,Acta}.

Now let us discuss the generalization of the fermion-loop scheme.
Remember that it provides both the NLO precision and the gauge
invariance. As a matter of fact the generalization consists in the
instruction on how to use the already known results of the former
calculations. Indeed, the first term in formula (34) and the last
two terms are already known \cite{Fermion-loop,On-shell,W-decay}.
The remaining second term is actually known too, since $\Im
\Sigma_2 (0)$ in it may be replaced by a one-loop correction to
the width of W-boson divided by its mass (see Property~3 and
discussion thereof).

The last two terms in (34) present exhaustive description of the
bosonic corrections, which formerly were a ``stumbling block'' for
gauge invariance. In the proposed generalization they are located
in the $S$-matrices squared, with the one $S$-matrix is for the
on-shell production of unstable particles and the other ones are
for their decays. The key observation that leads us to this simple
result is the absence in the NLO approximation of non-factorizable
corrections of Fig.~\ref{fig:2} (see Sect.~\ref{sec:7}). This
property demonstrates one of the differences of the approach
offered with the well known double-pole approximation (DPA)
\cite{Doubl-Pole}.

In fact there are more serious differences, too. The most obvious
one is as follows. By definition, DPA does not take into account
the single-pole contributions, while the fermion-loop scheme, and
consequently our generalization, does. The mentioned missed
contributions are of order $\Gamma_W/M_W$, which amounts to
2--3\%. This is too much compared with the current accuracy of
LEP2 results. In view of this problem, DPA usually is applied for
the calculation of the radiative correction only, but not of the
Born term. The Born term is considered in the framework of the
conventional PT with ``by-hand'' substitution of Breit--Wigner's
propagators instead of the free propagators for unstable particles
\cite{Doubl-Pole}. However, the latter operation violates the
gauge invariance and leads to some error, too, and a certain
estimate of this error does not exist. All that it is possible to
say is that this error most likely is somewhat less than the shift
of the amplitude caused by this substitution, but it is not clear
how far less. Nevertheless, the latter quantity again is
$O(\Gamma_W/M_W)$. So, the discussed approximation of the Born
term cannot be considered satisfactory.

Another serious difficulty of DPA is its inapplicability in the
vicinity of phase-space boundaries. This effect arises as a result
of the ``mapping'' procedure in the phase space or the ``analytic
continuation'' of the constant residues at the double pole in the
amplitude \cite{Doubl-Pole}. First of all, it manifests itself
near the threshold region, where the DPA uncertainties are blowing
up. Another evident restriction of applicability of DPA is the
region lying far off from resonance where the pole-scheme
expansion cannot be viewed as an effective expansion in powers of
$\Gamma_W/M_W$. Remember that the proposed generalization of the
fermion-loop scheme does not need the mapping procedure and does
not use the pole expansion. Consequently, it should be free from
these difficulties.

In conclusion, let us discuss two problems that arise in the
modified PT approach. In fact, the topic is about a comparison of
the results of the modified PT with Breit--Wigner's
parameterization of unstable particles. The first problem concerns
the definition of the ``physical'' mass and the width of unstable
particles. In this connection it should be noted, first of all,
that both these quantities are pseudo-observables that are to be
determined on the basis of realistic observables
\cite{B-P,Bardin}. Next, let us note that in the framework of the
modified PT both these quantities are secondary ones, which should
be determined on the basis of primary objects, such as the
renormalized Lagrangian mass, coupling constant, etc. Apparently,
the most radical way consists of the identification of both these
quantities with the position of the pole in the complex plane of
the full unstable propagator (in the spirit of the pole scheme).
This procedure is equivalent to finding a solution to the equation
$M^2\!-\!s_p\!-\!\Sigma (s_p)=0$ with $s_p=M_p^2\!-\!\i\kern 1pt
M_p\Gamma _p$. This operation may be done perturbatively
\cite{Sirlin}.

The second problem concerns the presence of the delta-functions
and VP's in the results of the complete AO expansion. (The
actuality of this problem is not so high while making use of
incomplete AO expansions as employed in this paper.) The problem
may be designated as an illusory discrepancy between the presence
of these singular functions in the amplitude squared and the
notion of a continuity of the physical results as functions of the
kinematic variables. This problem has been stated and discussed in
\cite{Base}. The essence of the solution is reduced to the
observation that actually there is not a direct identity between
the squared amplitude (that is usually calculated) and the genuine
probability (that is observed). In fact, a formal expression for
the probability, calculated on the basis of the unitarity
diagrams, should necessarily be subjected to the operation of
\emph{integration}, or ``\emph{smearing}'' with some weight
function before it becomes an observable quantity.\footnote{ It is
worth noticing, that with absence of ``smearing'' there is no
problem of non-integrable singularities in the phase space. In the
latter case all calculations actually are carried out in the
framework of the conventional PT.} In processes with light and
charged initial and/or final states such a smearing function is,
first of all, the flux function that extracts and exponentiates
the infrared and collinear singularities from the cross-section
(see formula (32)).

In the case of W- and/or Z-production the smearing is large
enough: on some distance from threshold its final effect is about
ten or more percent and it covers an area which is certainly
greater than the unstable-particle width (see formulas (32) and
(33)). Consequently, after the convolution the contributions of
the delta-functions and VP's become strongly distributed over a
wide area. Mathematically, this fact means that the smearing may
be considered as an effect of order $O(1)$, which is required in
the AO. Therefore, even in the case of ideal energy resolution in
a given experiment, the presence of the singular functions in the
results of the AO expansion becomes invisible, at least within the
given precision of the description. The above conclusion should
hold both for the total and differential cross-sections, since the
differential cross-sections are determined through the convolution
procedure, too \cite{LEP2}.

\bigskip

\noindent \emph{Acknowledgements.} The author is grateful to D.Yu.
Bardin, L.V. Kalinovskaya, and F.V. Tkachov for discussions of
various problems connected with applications of the electroweak
theory. Special thanks to F.V. Tkachov for the invitation to study
the problems of instability and consultation on the results of his
work \cite{Base}. The author would also like to thank A.A. Slavnov
for the valuable note about WI in presence of a photon mass, and
E.E. Boos for the indication of the importance of the
single-resonant subprocesses in W-pair production. It is a
pleasure to thank B.A. Arbuzov, V.E. Rochev and S.R. Slabospitsky
for support and useful discussions. This work was supported in
part by the Russian Foundation for Basic Research, grant No.
99-02-18365.


\begin{thebibliography}{99}
\bibitem{LEP2}
Report CERN 96-01, Physics at LEP2, eds. G.Altarelli, T.Sjostrand
and F.Zwirner.
\bibitem{Veltman}
M.Veltman, Physica \textbf{29} (1963) 186.
\bibitem{LEP2-1}
W.Beenakker et al., in \cite{LEP2}, p.79.
\bibitem{NuovoCim}
G.Montagna, O.Nicrosini and F.Piccinini, Rev.Nouvo Cim.
\textbf{21} (1998) 1.
\bibitem{B-P}
D.Bardin and G.Passarino, The standard model in the making.
Precision study of the electroweak interactions (Oxford Science
Pub., Clarendon Press, Oxford, 1999)
\bibitem{Stuart}
R.G.Stuart, Phys.Lett. \textbf{B262} (1991) 113.
\bibitem{Doubl-Pole}
W.Beenakker, F.Berends and A.Chapovsky, Nucl.Phys. \textbf{B548}
(1999) 3; A.Denner, S.Dittmaier, M.Roth and D.Wackeroth,
Nucl.Phys. \textbf{B560} (1999) 33.
\bibitem{Argyres}
E.N.Argyres et al., Phys. Lett. \textbf{B358} (1995) 339.
\bibitem{Fermion-loop}
W.Beenakker et al., Nucl.Phys. \textbf{B500} (1997) 255.
\bibitem{Background}
A.Denner and S.Dittmaier, Phys.Rev. \textbf{D54} (1996) 4499.
\bibitem{Dittmaier}
S.Dittmaier, Theoretical aspects of W physics, in Proc. HEP 97
(Jerusalem), eds. D.Lellouch (Springer-Verlag, 1999) p. 709;
Radiative corrections to W-pair production in $e^+e^-$
annihilation, in Proceedings RADCOR 98, Barcelona 1998, Radiative
corrections: Application of QFT to phenomenology, p.513.
\bibitem{Acta}
W.Beenakker and A.Denner, Acta Phys.Polon \textbf{B29} (98) 2821.
\bibitem{Base}
F.V.Tkachov, Perturbation theory with unstable fundamental fields,
hep-ph/9802307.
\bibitem{Schwartz}
L.Schwartz, Theorie des Distributions. I, II, Paris, 1950-51.
\bibitem{G-Sh}
I.M.Gelfand and G.E.Shilov, Generalized Functions (Academic Press,
1968)
\bibitem{IJMP}
F.V.Tkachov, Int.J.Mod.Phys. \textbf{A8} (1993) 2047.
\bibitem{PNP}
F.V.Tkachov, Phys.Part.Nucl. \textbf{25} (1994) 649.
\bibitem{Homo}
F.V.Tkachov, Phys.Lett. \textbf{B412} (1997) 350.
\bibitem{R}
N.N.Bogolyubov, Doklady USSR Acad. Sci. \textbf{82} (1952) 217.
\bibitem{B-Sh}
N.N.Bogolyubov and D.V.Shirkov, Introduction to the theory of
quantized fields (Fizmatgiz. Moscow 1957) (English transl.: Wiley,
1980, 3rd ed.)
\bibitem{Sirlin}
A.Sirlin, Phys.Rev.Lett. \textbf{67} (1991) 2127; Phys.Lett.
\textbf{B267} (1991) 240; R.G.Stuart, Phys.Rev.Lett. \textbf{70}
(1993) 3193.
\bibitem{Bardin}
D.Bardin, Field Theory and Standard Model. -- In: 1999 European
School of High Energy Physics. CERN Yellow Report, 2000-007. p.1.
\bibitem{Slavnov}
A.A.Slavnov, Phys.Lett. \textbf{B98} (1981) 57.
\bibitem{On-shell}
M.B\"{o}hm et al., Nucl.Phys. \textbf{B304} (1988) 463;
J.Fleischer, F.Jegerlehner and M.Zralek, Z.Phys. \textbf{C42}
(1989) 409.
\bibitem{W-decay}
D.Yu.Bardin, S.Riemann and T.Riemann, Z.Phys. \textbf{C32} (1986)
121; A.Denner and T.Sack, Z.Phys. \textbf{C46} (1990) 653.

\end{thebibliography}
\end{document}